\documentclass[prb,twocolumn,nofootinbib]{revtex4-1}
\usepackage[a4paper,top=2.50cm,bottom=2.50cm,left=2.50cm,right=2.50cm]{geometry}
\usepackage{epsfig}
\usepackage{rotate}
\usepackage{rotating}
\usepackage[abs]{overpic}
\usepackage{graphicx}
\usepackage{color}
\usepackage{ifpdf}
\usepackage{epstopdf}
\DeclareGraphicsExtensions{.eps, .pdf, .jpg, .tif}
\usepackage[]{hyperref}
\usepackage{amsmath}
\usepackage{amssymb}
\usepackage{bm}

\newcommand{\vpartial}{\mbox{\boldmath$\partial$}}
\def\ket#1{\mbox{$\displaystyle\vert\,#1\,\rangle$}}

\def\vj{{\bf j}} 
\def\vn{{\bf n}} 
\def\vq{{\bf q}} 
\def\vv{{\bf v}} 
 
\def\vz{{\bf z}} 
\def\vy{{\bf y}} 
\def\vx{{\bf x}} 
\def\vc{{\bf c}} 
\def\vd{{\bf d}} 
\def\vk{{\bf k}} 
\def\vS{{\bf S}} 
\def\vH{{\bf H}} 
\def\vB{{\bf B}} 
\def\vA{{\bf A}} 
\def\vM{{\bf M}} 
\def\vD{{\bf D}} 
\newcommand{\veta}{\mbox{\boldmath$\eta$}}
\newcommand{\ie}{{\it i.e.\ }}
\newcommand{\et}{{\it et al.\ }}
\newcommand{\eg}{{\it e.g.\ }}
\newcommand{\be}{\begin{equation}}
\newcommand{\ee}{\end{equation}}
\newcommand{\ber}{\begin{eqnarray}}
\newcommand{\eer}{\end{eqnarray}}
\newcommand{\vsigma}{\vec\sigma}
\newcommand{\vDelta}{\vec{\Delta}}
\newcommand{\vcY}{\vec{\cal Y}}
\def\ns{\negthickspace}

\begin{document}
\title{
\vspace*{-1.5cm}
\hspace*{-5.0cm}
{\sl\small Published in Adv. Phys. {\bf 43}, 113 (1994).
\href{http://dx.doi.org/10.1080/00018739400101475}
     {[doi:10.1080/00018739400101475]}
}
\\
\vspace*{1.0cm}
The Order Parameter for the Superconducting Phases of UPt$_3$
}
\author{J.A. Sauls}
\affiliation{Nordita, Blegdamsvej 17, DK-2100 Copenhagen \O, Denmark}
\altaffiliation{permanant address: Department of Physics \& Astronomy,
                Northwestern University, Evanston, IL 60208 USA}
\date{submitted to {\it Advances in Physics} - October 24, 1993}
\begin{abstract}
I review the principal theories that have been proposed for the superconducting phases of UPt$_3$. The detailed H-T phase diagram places constraints on any theory for the multiple superconducting phases. Much attention has been given to the Ginzburg-Landau (GL) region of the phase diagram where the phase boundaries of three phases appear to meet at a tetracritical point. It has been argued that the existence of a tetracritical point for all field orientations eliminates the two-dimensional (2D) orbital representations coupled to a symmetry breaking field (SBF) as  viable theory of these phases, and favors either (i) a theory based on two primary order parameters belonging to different irreducible representations that are accidentally degenerate [Chen and Garg, Phys.  Rev. Lett. 70, 1689 (1993)], or (ii) a spin-triplet, orbital one-dimensional (1D) representation with no spin-orbit coupling in the pairing channel [Machida and Ozaki, Phys. Rev. Lett. 66, 3293 (1991)].  I comment on the limitations of the models proposed so far for the superconducting phases of UPt$_3$. I also find that a theory in which the order parameter belongs to an orbital 2D representation coupled to a SBF is a viable model for the phases of UPt$_3$, based on the existing body of experimental data. Specifically, I show that (1) the existing phase diagram (including an apparent tetracritical point for all field orientations), (2) the anisotropy of the upper critical field over the full temperature range, (3) the correlation between superconductivity and basal plane antiferromagnetism and (4) low-temperature power laws in the transport and thermodynamic properties can be explained qualitatively, and in many respects quantitatively, by an odd-parity, E$_{2u}$ order parameter with a pair spin projection of zero along the ${\vc}$-axis. The coupling of an AFM moment to the superconducting order parameter acts as a symmetry breaking field (SBF) which is responsible for the apparent tetracritical point, in addition to the zero-field double transition. The new results presented here for the E$_{2u}$ representation are based on an analysis of the material parameters calculated within BCS theory for the 2D representations, and a refinement of the SBF model of Hess, {\et} [J. Phys. Condens. Matter, 1, 8135 (1989)]. I also discuss possible experiments to test the symmetry of the order parameter.\footnote{Presented at the symposium on {\it Kondo Lattices and Heavy Fermions} at the American Physical Society Meeting held in Seattle, Washington, March 1993.  Published in {\it Advances in Physics}, \underline{43}, pp.  113-141 (1994).}
\end{abstract}
\maketitle
\thispagestyle{plain}
\setcounter{page}{1}

\vspace*{-13mm}
\subsection*{Introduction}
\vspace*{-3mm}

It is almost a clich\'{e} to say that many of the heavy fermion superconductors are thought to represent a novel form of superconductivity. But, in spite of considerable progress experimentally and theoretically we have not yet firmly identified the order parameter, even for the prime candidate UPt$_3$. To date liquid $^3$He is the only material that we are certain exhibits unconventional BCS pairing. Superfluid $^3$He was discovered in 1972, and within three years the identification of the phases with spin-triplet, p-wave order parameters was essentially complete.\cite{leg75,vollhardt90} It is nine years since superconductivity in UPt$_3$ was discovered,\cite{ste84} yet there is no conscensus about the identification of the order parameter. Of course the heavy fermion materials are much more complex, while $^3$He is perhaps the purest elemental substance known. The issue of material quality is critical; pairing correlations in unconventional superconductors are known to be sensitive to scattering from defects. Indeed the discovery of the multiple superconducting transitions in UPt$_3$ was made on single crystals of high quality and purity.\cite{mul87,qia87,sch89,fis89,has89} Five superconducting phases have now been identified experimentally - two Meissner phases and three flux phases.\cite{bru90,ade90,bul93} The observations of basal plane AFM order,\cite{aep88,fri88} and its correlation with the zero-field superconducting transition\cite{tra91,hay92} certainly identify the phases of UPt$_3$ as one of the remarkable examples of complex symmetry breaking in any material.

In this article I examine the principal theories that have been proposed for the superconducting phases of UPt$_3$. The detailed H-T phase diagram that is now available places stringent constraints on theories for the multiple superconducting phases. Much attention has been given to the Ginzburg-Landau (GL) region of the phase diagram where the phase boundaries of three phases appear to meet at a tetracritical point. Machida and Ozaki\cite{mac91} and Chen and Garg\cite{che93} have argued that the existence of a tetracritical point for all field orientations eliminates the two-dimensional (2D) orbital representations coupled to a symmetry breaking field (SBF)\cite{hes89,mac89} as viable theory of these phases, and favors either (i) a theory based on two primary order parameters belonging to different irreducible representations that are accidentally degenerate,\cite{che93} or (ii) a spin-triplet, orbital one-dimensional (1D) representation with no spin-orbit coupling in the pairing channel.\cite{mac91} I discuss the models proposed so far for the superconducting phases of UPt$_3$.

I argue that the 2D model coupled to a SBF is a viable model for the phases of UPt$_3$ based on the existing experimental data.  However, the existence of an apparent tetracritical point for all field orientations restricts the order parameter to the E$_{2}$ representation. Furthermore, (1) the existing H-T phase diagram, (2) the anisotropy of the upper critical field over the full temperature range, (3) the correlation between superconductivity and basal plane antiferromagnetism and (4) low-temperature power laws in the transport and thermodynamic properties can be explained qualitatively and in many respects quantitatively by an odd-parity, E$_{2u}$ order parameter with the spin quantization axis aligned such that ${\vS}_{pair}\cdot{\vc}=0$. The coupling of an AFM moment to the superconducting order parameter is a SBF, which in this theory is responsible for the apparent tetracritical point as well as the double transition in zero field. The new results presented here for the E$_{2u}$ representation are based on an analysis of the material parameters within BCS theory for the 2D representations and a refinement of the SBF model of Hess, {\et}\cite{hes89} I conclude with a discussion of possible experimental tests of the residual symmetry and broken symmetries in the ordered phases.

\vspace*{-3mm}
\subsection*{Multiple Superconducting Phases of UPt$_3$}
\vspace*{-3mm}

Considerable evidence in support of an unconventional superconducting state\footnote{This term is used here to mean a superconductor in which the order parameter spontaneously breaks one or more symmetries in combination with U(1) gauge symmetry.} in the heavy fermion materials has accumulated from specific heat,\cite{sul86,fis89} upper critical field\cite{shi86b,tai88,vor92} and various transport measurements,\cite{bis84,mul86,shi86a,sul86,gro88,shi90,sig92} all of which show anomalous properties compared to those of conventional superconductors. However, the strongest evidence for unconventional superconductivity comes from the multiple superconducting phases of UPt$_3$.
 
\begin{figure}
\includegraphics[width=\columnwidth]{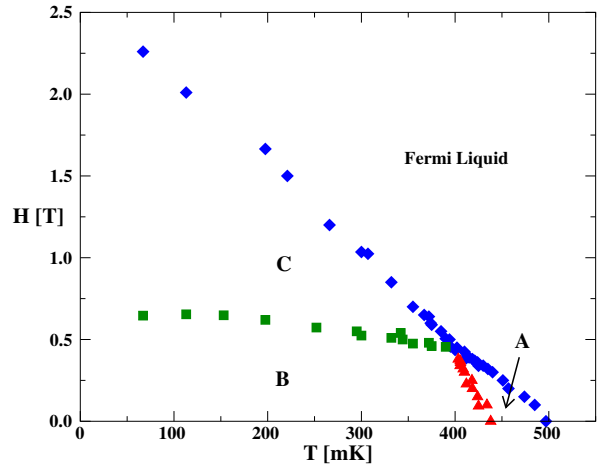}
\caption{
Phase diagram for $\vH \perp {\vc}$ from Ref. (\onlinecite{ade90}) showing three superconducting phases.
Points represent transitions between phases determined from sound velocity anomalies. 
The lower critical field lines separating the Meissner states for the A and B phases are not shown.
}
\label{fig:HT_phasediagram}
\end{figure}

The H-T phase diagram of superconducting UPt$_3$ is unique; there are two superconducting phases in zero field,\cite{fis89,has89} and three vortex phases.\cite{mul87,qia87,sch89,kle89} This phase diagram has been mapped out using ultrasound velocity measurements,\cite{bru90,ade90,bul93} and with dilatometry.\cite{van93a} The sound velocity, which at low frequencies ($\omega\ll\tau^{-1}$) is a thermodynamic property, shows an anomaly anologous to the specific heat discontinuity at a second-order mean-field phase transition. The sound velocity is proportional to the second derivative of the free energy with respect to strain; near a second-order mean-field phase transition the velocity exhibits a discontinuity that is proportional to the discontinuity in the heat capacity, $\frac{\Delta v}{v}\propto\frac{\partial^2 F}{\partial\epsilon^2} \propto -\frac{\Delta C}{C_N}\,\left(\frac{\partial T_c}{\partial \epsilon}\right)^2$.\cite{tha91} Velocity anomalies of order $1 - 30$ ppm were measured by Bruls, {\et}\cite{bru90}, Adenwalla, {\et}\cite{ade90} and Bullock, {\et}\cite{bul93} and a phase diagram was 
constructed (Fig.~\ref{fig:HT_phasediagram}).
%
%
There are several important features of the phase diagram which I summarize from Ref. (\onlinecite{ade90}):
\begin{enumerate}
\item There are two zero-field superconducting phases with a difference in $T_c$ of
      $\frac{\Delta T_c}{T_c}\simeq 0.1$.
\item A change in slope of $H_{c2}(T)$ (a `kink' in $H_{c2}^{\perp}$) is observed 
      for ${\vH}\perp{\vc}$, but not for ${\vH}||{\vc}$.
\item There are three flux phases. The phase transition lines separating the flux phases 
      appear to meet at a tetracritical point for ${\vH}\perp{\vc}$, ${\vH}||{\vc}$ and 
      $\cos^{-1}({\vH}\cdot{\vc})=45^o$, although the case for a tetracritical point is 
      strongest for ${\vH}\perp{\vc}$. The resolution of the phase transition lines 
      near the tetracritical point is $\approx 13\,mK$.
\end{enumerate}
 
All of the models to date start from the basic assumption that the phases are all related to an equal-time pairing amplitude,
\be
\Delta_{\alpha\beta}({\vk}_f)\sim
\left<a_{{\vk}_f\alpha}a_{-{\vk}_f\beta}\right>
\,,
\ee
where $\alpha,\beta$ refer to the pseudo-spin labels of the quasiparticles. In the heavy fermion materials it is generally assumed that the spin-orbit interaction is strong;\cite{and84a,vol85,lee86} thus, the labels characterizing the quasiparticle states near the Fermi level are not eigenvalues of the spin operator for electrons. Nevertheless, in zero-field the Kramers degeneracy guarantees that each ${\vk}$ state is two-fold degenerate, and thus, may be labeled by a pseudo-spin quantum number $\alpha$, which can take on two possible values.  Furthermore, the degeneracy of each ${\vk}$-state is lifted by a magnetic field, which is described by a Zeeman energy that couples the magnetic field to the pseudo-spin with an effective moment that in general depends on the orientation of the magnetic field relative to crystal coordinates, and possibly the wavevector ${\vk}$.  As I discuss below, the Zeeman energy plays an important role in the superconducting state.

Fermion statistics of the quasiparticles requires the pair amplitude to obey the anti-symmetry condition,
\be
\Delta_{\alpha\beta}({\vk}_f)=-\Delta_{\beta\alpha}(-{\vk}_f)
\,.
\ee
Essentially all of the heavy fermion superconductors, including UPt$_3$, have inversion symmetry. This has an important consequence for the allowed classes of superconductivity;\cite{and84a,vol84} the pairing interaction that drives the superconducting transition necessarily decomposes into even- and odd-parity sectors. Thus, $\Delta_{\alpha\beta}({\vk}_f)$ for any theory based on a single primary order parameter necessarily has even or odd parity, and therefore, is pseudo-spin `singlet' or pseudo-spin `triplet' (I drop the `pseudo' hereafter.),
\ber
\Delta_{\alpha\beta}({\vk}_f)&=&\Delta({\vk}_f)\,(i\sigma_y)_{\alpha\beta}
\qquad {\rm singlet}\, (S=0)
\,,
\\
\Delta_{\alpha\beta}({\vk}_f)&=&{\vDelta}({\vk}_f)\cdot
(i{\vsigma}\sigma_y)_{\alpha\beta}
\quad {\rm triplet}\, (S=1)
\,,
\eer
with $\Delta({\vk}_f)=\Delta(-{\vk}_f)$ (even parity) and ${\vDelta}({\vk}_f)=-{\vDelta}(-{\vk}_f)$ 
(odd-parity).\footnote{Unconventional pairing states which have no equal-time average, {\ie} `odd-frequency' pairing states, obey a more general anti-symmetry condition which includes the imaginary time coordinate. Such states have the parity and total spin quantum numbers interchanged. See Berezinskii,\cite{ber74} and for a more recent discussion of these states, Balatsky and Abrahams.\cite{bal92}} 
Furthermore, the pairing interaction separates into a sum over invariant bilinear products of basis functions for each irreducible representation $\Gamma$ of the crystal point group, for both even- (${\cal Y}_{\Gamma,i}({\vk}_f)$) and odd-parity (${\vcY}_{\Gamma,i}({\vk}_f)$) sectors. A complete tabulation of the basis functions for the symmetry groups of the heavy fermion superconductors is given in Ref. (\onlinecite{yip93c}), and a list of the irreducible representations and representative basis functions for the group $D_{6h}$, appropriate for UPt$_3$ with strong spin-orbit coupling, is given in Table \ref{tab_basis}. The general form of the order parameter is then,
\ber
\Delta({\vk}_f)
&=&
\sum_{\Gamma}^{even}\sum_i^{d_{\Gamma}}\,\eta_{i}^{(\Gamma)}\,{\cal Y}_{\Gamma,i}({\vk}_f)
\,,
\\
{\vDelta}({\vk}_f)
&=&
\sum_{\Gamma}^{odd}\sum_i^{d_{\Gamma}}\,\eta_{i}^{(\Gamma)}\,{\vcY}_{\Gamma,i}({\vk}_f)
\,.
\eer

The actual realization of superconductivity is determined by the order parameter which minimizes the free energy. The Ginzburg-Landau (GL) theory is formulated in terms of a stationary free energy functional of the pair amplitude, and is constructed from basic symmetry considerations. The central assumptions are that the free energy functional can be expanded in powers of the order parameter and that the GL functional has the full symmetry of the normal state. The leading order terms in the GL functional are of the form,
\be
{\cal F}=\int\,d^3x\,
\sum_{\Gamma}\,\alpha_{\Gamma}(T)\,\sum_{i=1}^{d_{\Gamma}}
|\eta^{(\Gamma)}_{i}|^2 + ...
\,.
\ee
There is a single quadratic invariant for each irreducible representation. The coefficients $\alpha_{\Gamma}(T)$ are material parameters that depend on temperature and pressure. Above $T_c$ all the coefficients $\alpha_{\Gamma}(T)>0$. The instability to the superconducting state is then the point at which one of the coefficients vanishes, {\eg} $\alpha_{\Gamma^*}(T_c)=0$. Thus, near $T_c$ $\alpha_{\Gamma^*}(T)\simeq\alpha'(T-T_c)$ and $\alpha_{\Gamma}>0$ for $\Gamma\ne\Gamma^*$. At $T_c$ the system is unstable to the development of all the amplitudes $\{\eta_i^{(\Gamma^*)}\}$, however, the higher order terms in the GL functional which stabilize the system, also select the ground state order parameter from the manifold of degenerate states at $T_c$. In most superconductors the instability is in the even-parity, $A_{1g}$ channel. This is conventional superconductivity in which only gauge symmetry is spontaneously broken. An instability in any other channel is a particular realization of unconventional superconductivity.

\begin{table}\squeezetable
\begin{tabular}{|c|c|c|c|}
\hline
& Even parity & & Odd parity 
\\ 
\hline
A$_{1g}$ & 1 &
A$_{1u}$ & ${\vz}\,k_z$  
\\
\hline 
A$_{2g}$ & $Im(k_x+ik_y)^6$ &
A$_{2u}$ & ${\vz}\,k_z Im(k_x+ik_y)^6$  
\\
\hline 
B$_{1g}$ & $k_z\,Im(k_x+ik_y)^3$ &
B$_{1u}$ & ${\vz}\,Im(k_x+ik_y)^3$ 
\\
\hline 
B$_{2g}$ & $k_z\,Re(k_x+ik_y)^3$ &
B$_{2u}$ & ${\vz}\,Re(k_x+ik_y)^3$  
\\
\hline 
E$_{1g}$ & $k_z\left(\begin{array}{c}k_x \\ k_y \end{array}\right)$ &
E$_{1u}$ & ${\vz}\left(\begin{array}{c}k_x \\ k_y
 \end{array}\right)$ 
\\
\hline 
E$_{2g}$ & $\left(\begin{array}{c}k_x^2-k_y^2\\2k_xk_y
 \end{array}\right)$ &
E$_{2u}$ & ${\vz}\,k_z
 \left(\begin{array}{c}k_x^2-k_y^2\\2k_xk_y \end{array}\right)$
\\
\hline 
\end{tabular}
\caption{Basis functions for $D_{6h}$}
\label{tab_basis}
\end{table}

There are two basic types of models that have been proposed to explain the phase diagram of UPt$_3$: (i) theories based on a {\it single} primary order parameter belonging to a higher dimensional representation of the $D_{6h}$ symmetry group of the normal state, and (ii) theories based on {\it two} primary order parameters belonging to different irreducible representations of $D_{6h}$ which are nearly degenerate.

\vspace*{-3mm}
\subsection*{Two-dimensional models with symmetry breaking}
\vspace*{-3mm}
 
The first class includes the theory proposed by Hess, {\et}\cite{hes89} and Machida and Ozaki.\cite{mac89} This model assumes that the primary order parameter belongs to one of the four possible two-dimensional representations. The Ginzburg-Landau functional is constructed from the amplitudes that parametrize $\Delta_{\alpha\beta}({\vk}_f)$, {\it e.g.} if $\Delta$ belongs to the E$_{2u}$ representation listed in Table \ref{tab_basis},
\be
{\vDelta}({\vk}_f) =
{\vz}\left(\eta_1 \,k_z(k_x^2-k_y^2)+\eta_2 \,2k_zk_xk_y\right)
\,,
\ee
the GL order parameter is then a complex two-component vector, ${\veta}=(\eta_1,\eta_2)$, transforming according to the E$_2$ representation. In the case of strong spin-orbit coupling, the terms in the GL functional must be invariant under the symmetry group, $G=D_{6h}\times{\cal T}\times U(1)$, of point rotations, time-reversal and gauge transformations. The form of ${\cal F}$ is governed by the linearly independent invariants that can be constructed from fourth-order products, $\sum\,b_{ijkl}\,\eta_i\eta_j\eta_k^*\eta_l^*$, and second-order gradient terms, $\sum\,\kappa_{ijkl}(D_i\eta_j)(D_k\eta_l)^*$. For the 2D representations there are two independent fourth-order invariants and four independent second-order gradients; the GL functional has the general form,\cite{vol84,gor87,joy88,hes89,mac89,tok90,joy90,sig91}

\ber\label{free_energy}
{\cal F}
&\ns=\ns&
\int\ns d^{3}x\Big\lbrace\,\alpha(T){\veta}\cdot{\veta}^{*}
+
\beta_{1}\left({{\veta}\cdot{\veta}^{*}}\right)^{2}
+
\beta_{2}\left\vert{{\veta}\cdot{\veta}}\right\vert^{2}
\nonumber\\ 
&+& 
\kappa_{1}\left({D_{i}\eta_{j}}\right)\left({D_{i}\eta_{j}}\right)^{*}
+
\kappa_{2}\left({D_{i}\eta_{i}}\right)\left({D_{j}\eta_{j}}\right)^{*}
\nonumber\\ 
&+&
\kappa_{3}
\left({D_{i}\eta_{j}}\right)\left({D_{j}\eta_{i}}\right)^{*}
+
\kappa_{4}
\left({D_{z}\eta_{j}}\right)\left({D_{z}\eta_{j}}\right)^{*} 
\nonumber\\ 
&+&
\frac{1}{8\pi}|{\vpartial\times\vA}|^2
\Big\rbrace 
\,,
\eer
where ${\cal F}\left[{{\veta},{\vA\;}}\right]$, at its minimum, is the difference between the superconducting and normal-state free energies, $[\alpha (T),\beta_1,\beta_2,\kappa_1,\kappa_2,\kappa_3,\kappa_4]$ are material parameters that can be calculated from microscopic theory, or be determined from comparision with experiment, and 
$\vpartial\times\vA=\vB$ is the magnetic field. The gauge-invariant derivatives are denoted by $D_i=\partial_i+i\frac{2e}{\hbar c}A_i$.

The equilibrium order parameter and current distribution are determined by the stationarity conditions of the GL functional with respect to variations of the order parameter and the vector potential. These conditions yield the GL differential equations,\cite{tok90}
\begin{widetext}
\be
\kappa_{123}D_{x}^{2}\eta_{1}
+\kappa_{1}D_{y}^{2}\eta_{1}
+\kappa_{4}D_{z}^{2}\eta_{1}
+(\kappa_{2}D_{x}D_{y}+\kappa_{3}D_{y}D_{x})\eta_{2}
+2\beta_{1}\left({{\veta}\cdot{\veta}^{*}}\right)\eta_{1}
+2\beta_{2}\left({{\veta}\cdot{\veta}}\right)\eta^{*}_{1}
= \alpha\,\eta_{1}
\,,
\ee
\be
\kappa_{1}D_{x}^{2}\eta_{2}
+\kappa_{123}D_{y}^{2}\eta_{2}
+\kappa_{4}D_{z}^{2}\eta_{2}
+(\kappa_{2}D_{y}D_{x}+\kappa_{3}D_{x}D_{y})\eta_{1}
+2\beta_{1}\left({{\veta}\cdot{\veta}^{*}}\right)\eta_{2}
+2\beta_{2}\left({{\veta}\cdot{\veta}}\right)\eta^{*}_{2}
= \alpha\,\eta_{2}
\,,
\ee
and the Maxwell equation,
\be
(\vpartial\times{\vB})_{i}=
-{{16\pi e}\over{\hbar c}}Im
[\kappa_{1}\,\eta_{j}\left({D_{\perp,i}\eta_{j}}\right)^{*}+
\kappa_{2}\,\eta_{i}\left({D_{\perp,j}\eta_{j}}\right)^{*}+
\kappa_{3}\,\eta_{j}\left({D_{\perp,j}\eta_{i}}\right)^{*}+
\,\kappa_{4}\,\delta_{iz}\eta_{j}\left({D_{z}\eta_{j}}
\right)^{*}]
\,,
\ee
\end{widetext}
which are the basis for studies of the H-T phase diagram, vortices and related magnetic properties.\cite{sch89,sig89,tok90,tok90a,pal90,bar91,joy91,mel92,pal92,pal92a} Note that $\kappa_{ijk ...}=\kappa_i + \kappa_j + \kappa_k + \,...$.

There are two possible homogeneous equilibrium states depending on the sign of $\beta_2$. For $-\beta_1<\beta_2<0$ the equilibrium order parameter, ${\veta}=\eta_0{\vx}$ (or any of the six degenerate states obtained by rotation), breaks rotational symmetry in the basal plane, but preserves time-reversal symmetry. However, for $\beta_2>0$ the order parameter retains the full rotational symmetry (provided each rotation is combined with an appropriately chosen gauge transformation), but spontaneously breaks time-reversal symmetry. The equilibrium state is doubly-degenerate with an order parameter of the form ${\veta}_{+}=(\eta_0/\sqrt{2})({\vx}+i{\vy})$ [or ${\veta}_{-}={\veta}_{+}^{*}$], where $\eta_0=\sqrt{|\alpha|\over 2\beta_1}$. The broken time-reversal symmetry of the two solutions, ${\veta}_{\pm}$, is exhibited by the two possible orientations of the internal orbital angular momentum,
\be
{\vM}_{orb}=(\kappa_2-\kappa_3)\left(\frac{2e}{\hbar c}\right)Im\left({{\veta}\times{\veta}^{\,*}}\right)\sim\pm{\vz}\,,
\ee
or spontaneous magnetic moment of the Cooper pairs. The presence of this term in the GL functional is not transparent from eq.(\ref{free_energy}). However, the gradient terms in the GL functional can be rewritten in the following form,
\begin{widetext}
\ber
{\cal F}_{\mbox{\tiny grad}}
&=&
\int d^3x\,
\Big\{
\kappa_1\,[|{\vD}_{\perp}\eta_1|^2+|{\vD}_{\perp}\eta_2|^2]+
+
\kappa_4\, [|D_z\eta_1|^2+|D_z\eta_2|^2]
+
\kappa_{23}(|D_x\eta_1|^2+|D_y\eta_2|^2)
\nonumber\\
&+&
\frac{1}{2}\kappa_{23}
\left[(D_x\eta_1)(D_y\eta_2)^* + (D_x\eta_2)(D_y\eta_1)^* +c.c.\right]
+
(\kappa_2-\kappa_3)\left[\left(\frac{2e}{\hbar c}\right)
(i{\veta}\times{\veta}^*)\cdot({\vpartial}\times{\vA})\right]
\Big\}
\,,\quad
\label{f_grad}
\eer
\end{widetext}
revealing the coupling of the orbital moment of the pairs to the magnetic field. Since the coupling
of the order parameter to a magnetic field is primarily diamagnetic (for $T\simeq T_c$), the orbital moment is difficult to observe because of Meissner screening.\footnote{There are, however, subtleties associated with the orbital moment. The orbital Zeeman energy is not locally positive. Thus, if the coupling of the orbital moment to the magnetic field is sufficiently strong the favored ground state, even in zero applied field, is a non-uniform current-carrying state.\cite{pal90,pal92} However, estimates of the GL parameters do not favor such a ground state in UPt$_3$.}
 
The case $\beta_2>0$ is relevant for the 2D models of the double transition of UPt$_3$. However, the 2D theory has only one phase transition in zero field, and by itself cannot explain the double transition. The small splitting of the double transition in UPt$_3$ ($\Delta T_c/T_c\simeq 0.1$) suggests the presence of a small symmetry breaking energy scale and an associated lifting of the degeneracy of the possible superconducting states belonging to the 2D representation. The second zero-field transition just below $T_c$ in UPt$_3$, as well as the anomalies observed in the upper and lower critical fields, have been explained in terms of a weak symmetry breaking field (SBF) that lowers the crystal symmetry from hexagonal to orthorhombic, and consequently reduces the 2D E$_2$ (or E$_1$) representation to two 1D representations with slightly different transition temperatures.\cite{hes89,mac89} The key point is that right at $T_c$ all phases of the 2D representation are degenerate, thus any SBF that couples second-order in ${\veta}$ and prefers a particular phase will dominate near $T_c$. At lower temperatures the SBF energy scale, $\Delta T_c$, is a small perturbation compared to the fourth order terms in the fully developed superconducting state and one recovers the results of the GL theory for the 2D representaion, albeit with small perturbations to the order parameter.
 
In UPt$_3$ there appears to be a natural candidate for a SBF;\cite{joy88} the AFM order in the basal plane reported by Aeppli, {\et}\cite{aep88}, Frings, {\et}.\cite{fri88} and Hayden, {\et}\cite{hay92} The lowest order invariant that can be constructed from the in-plane AFM order parameter, ${\vM}_s$, and the superconducting order parameter, ${\veta}$, is
\be
{\cal F_{{\rm SBF}}}\left[{{\veta}}\right]
=\epsilon\,M_s^2\,\int\limits d^{3}x
\,(|\eta_1|^2 - |\eta_2|^2)
\;,
\ee
where the coupling parameter $\epsilon M_s^2$ determines the magnitude of the splitting of the superconducting transition (we denote the first transition by $T_c$ and the lower superconducting transition by $T_{c*}$).\footnote{A SBF coupling of this form is allowed for a superconducting order parameter belonging to either an $E_1$ or $E_2$ representation. The non-symmetry-breaking invariant of the same order, $M_s^2(|\eta_1|^2 + |\eta_2|^2)$, is absorbed into $\alpha|{\veta}|^2$, which shifts $T_{c0}$.} 
The analysis of this GL theory, including the SBF, is given in Ref.(\onlinecite{hes89}); 
I summarize the main results below:
\begin{enumerate}
\item A double transition in zero field occurs only if $\beta_2>0$. The splitting of the transition temperature is $\Delta T_c\propto \epsilon\,M_s^2$.
\item The relative magnitudes of the heat capacity anomalies, $\Delta C_*/T_{c*} > \Delta C/T_c$, are consistent with the stability condition $\beta_2>0$ required in order for there to be two superconducting phases in the 2D model. Heat capacity measurements by several groups give $\beta_2/\beta_1 \simeq 0.2\,-\,0.5$.\cite{fis89,has89,vor92}
\item The low temperature phase ($T<T_{c*}$) has broken time-reversal symmetry, and is doubly degenerate: ${\veta}_{\pm}\sim (a(T),\pm i\,b(T))$, reflecting the two orientations of the internal angular momentum of the ground state.
\item The high temperature phase has broken rotational symmetry in the
basal plane, induced by the SBF. The basal plane anisotropy of the order parameter then leads to anisotropic current flow and consequently an orthorhombic vortex lattice for ${\vH}||{\vc}$. However, the orthorhombic distortion will be small if $\kappa_{23}\ll\kappa_1$ as I argue below.  \item The upper critical field exhibits a change in slope, a `kink' at high temperature for ${\vH}\perp{\vc}$, but not for ${\vH}||{\vc}$. The kink in $H_{c2}^{\perp}$ is isotropic in the basal plane provided the in-plane magnetic anisotropy energy is weak compared to the Zeeman energy acting on ${\vM}_s$, in which case ${\vM}_s$ rotates to maintain ${\vM}_s\perp{\vH}$. Recent magnetoresistance experiments support the interpretation of weak magnetic anisotropy energy in the basal plane.\cite{vor92}
\item Kinks in $H_{c1}$, for all field orientations, are predicted to occur at the second zero-field transition temperature, $T_{c*}$.  This has been confirmed by several groups.\cite{shi89,vin91,kne92,vor92} The increase in $H_{c1}\propto \eta^2$ at $T_{c*}$ is also a strong indication of the onset of a second superconducting order parameter.
\item Additional evidence for a SBF model of the double transition comes from pressure studies of the superconducting and AFM phase transitions. Heat capacity measurements by Trappmann, {\et}\cite{tra91} show that both zero-field transitions are suppressed under hydrostatic pressure, and that the double transition disappears at $p_{*}\simeq 4\,kbar$. Neutron scattering experiments reported by Hayden, {\et}\cite{hay92} show that AFM order disappears on the same pressure scale, at $p_c\simeq 3.2\,kbar$. The qualitative fact that $p_{*}>p_c$ is consistent with $\Delta T_c(p)\sim M_s^2(p,T_{c+}(p))$ from the SBF model and the pressure dependences of $T_c$, $M_s^2$ and $T_{afm}$.\footnote{While the decrease of $T_c(p)$ is clear from experiments, the pressure dependence of $T_{afm}$ is not. Trappmann, {\et}\cite{tra91} interpreted the neutron scattering data of Refs.(\onlinecite{aep88,tai90a}) to imply a linear decrease of $T_{afm}$ with increasing pressure. However, Hayden, {\et}\cite{hay92} observed no noticeable drop in $T_{afm}$ for pressures up to $p=2\,kbar$, but were unable to track the transition point to higher pressures.  Reasonable consistency between $\Delta T_c(p)\sim M_s^2(p,T_{c+}(p))$, $p_{*}>p_{cr}$, and the pressure derivatives is obtained for either extreme: (i) $T_{afm}\simeq {\rm constant}$ and $M_s^2(p,0)\propto(p_{cr}-p)$, or (ii) $T_{afm}(p)=T_{afm}(0)-\gamma p$ and $M_s^2(p,T)={\rm const}\,(T_{afm}(p)-T)$.}
This correlation between AFM order and the existence of the double transition is strong support for the 2D model and the SBF explanation of the double transition.  It is worth pointing out that the vanishing of the AFM order parameter in the P-T plane, $M_s^2(T,p_{cr}(T))=0$, also defines a nearly vertical second-order transition line of the superconducting order parameter that extends from the critical point at $\Delta T_c(p_*)=0$ towards $(p_{cr},0)$ in the P-T plane. The 2D order parameter recovers full $D_{6}[E]$ symmetry for $p>p_{cr}(T)$.
\end{enumerate}
 
The phase diagram determined by ultrasound velocity measurements indicates that the phase boundary lines meet at a tetracritical point for both ${\vH}||{\vc}$ and ${\vH}\perp{\vc}$.\footnote{The extrapolation of the phase diagram data to a tetracritical point is most convincing for ${\vH}\perp{\vc}$, but is plausible for the other field orientations as well. However, for the arguments that follow it does not matter whether or not the ciritcal region is a true tetracritical point because I do not discuss the symmetries of the vortex phases in the critical region. I shall use the term `apparent tetracritical point' to describe the critical region.} This has been argued to contradict the GL theory based on a 2D order parameter.\cite{mac91,che93} The difficulty arises from gradient terms in the free energy that couple the two components of the 2D order parameter.

Besides the orbital Zeeman term, the terms $\kappa_{23}[(D_x\eta_1)(D_y\eta_2)^* + (D_y\eta_1)(D_x\eta_2)^* +\,c.c.]$ in eq.(\ref{f_grad}), which couple $\eta_1$ and $\eta_2$, lead to `level repulsion' effects in the linearized GL differential equations that prevent the crossing of two $H_{c2}(T)$ curves corresponding to different eigenfunctions, {\ie} different superconducting phases.\footnote{I use the nomenclature of Garg.\cite{gar92} The determination of the critical fields lines is an eigenvalue problem, which in the GL region can be mapped onto the Schr{\"o}dinger equation for a charged particle with internal degrees of freedom (related to the symmetry class of the order parameter), moving in a spatially varying potential if there is a background vortex lattice.  In this language a tetracritical point is possible only if the critical fields are associated with eigenfunctions (order parameters) belonging to different quantum numbers of a conserved variable. In the absence of such a conserved quantum number `level repulsion' will prohibit the crossing of critical field lines, in which case there can be no true tetracritical point.\cite{luk91}}
The `level repulsion' vanishes for ${\vH}\perp{\vc}$ (with alignment of the SBF), but {\it not} for other field directions. This feature of the 2D model has spawned alternative theories, designed specifically to eliminate the `level repulsion' effect.\cite{mac91,luk91,zhi92,che93}

\vspace*{-3mm}
\subsection*{Odd-Parity and Weak Spin-Orbit Coupling Models}
\vspace*{-3mm}

Machida and Ozaki\cite{mac91} relax the assumption of strong spin-orbit
coupling in the pairing channel. Thus, the full symmetry group in their
model includes the continuous spin-rotation group,
$G=SO(3)_{spin}\times D_{6h}\times {\cal T} \times U(1)$. They preserve
the coupling of a SBF to superconductivity through the spin-triplet
components of the order parameter, and avoid the `level repulsion'
problem by choosing a 1D representation for the orbital component of
the order parameter. In their model
\be
{\vDelta}({\vk}_f) = {\vd}\,\,{\cal Y}({\vk}_f)
\,,
\ee
where the ${\vd}$ is a vector in spin space (in general complex)\footnote{The ${\vd}$ vector is the order parameter describing the spin correlations of the pairing state. If this vector is real, then ${\vd}$ is the direction along which the pair spin projection is zero, {\ie} ${\vd}\cdot{\vS}_{pair}=0$; if ${\vd}$ is complex, then the pairs are ordered into a spin-polarized state. For example, ${\vd}\sim {\vx}+i{\vy}$ implies ${\vz}\cdot{\vS}_{pair}=\hbar$.} and ${\cal Y}({\vk}_f)$ is the basis function for the orbital part of the order parameter. The coupling of the SBF to ${\vd}$ generates a sequence of transitions in which the $d_x$ and $d_y$ components are nonzero. Consider the GL free energy functional for this theory,\footnote{I use the ${\vd}$ vector notation to avoid confusion with the orbital 2D models; the GL functional eq. (\ref{GL_mac}) is equivalent to that of Machida and Ozaki.\cite{mac91} Note that ${\vd}$ is a 3D spin vector.}
\begin{widetext}
\ber\label{GL_mac}
{\cal F}\left[{{\vd},{\vA\;}}\right] 
&=& 
\int d^{3}x \;\Big\lbrace\,\alpha(T){\vd}\cdot{\vd}^{*}
+\epsilon\,|{\vM}_s\cdot{\vd}|^2
+\beta_{1}({\vd}\cdot{\vd}^{*})^{2}
+\beta_{2}|{\vd}\cdot{\vd}|^{2}
+g\,|{\vH}\cdot{\vd}|^2  
\nonumber\\ 
&&
\qquad\qquad
+\kappa_{\perp}\sum_{i}\,|{\vD}_{\perp}d_{i}|^2
+\kappa_{||}\sum_{i}\,|{\vD}_{z}d_{i}|^2
+\frac{1}{8\pi}|{\vpartial\times\vA}|^2
\Big\rbrace 
\,.
\eer
\end{widetext}

The SBF term is written in a compact form with ${\vM}_s=M_s {\vx}$. The first transition is to a phase with $d_x\ne 0$, followed by a second transition $T_{c2}=T_{c1} - {\cal O}(\epsilon M_s^2)$ in which both $d_y$ and $d_z$ nucleate with a phase $\pm \pi/2$ relative to that of $d_x$. As in the orbital 2D models, $\beta_2>0$ is required for a double transition and as a result time-reversal symmetry is broken below $T_c$, but now by the spin degrees of freedom. The term proportional to $|{\vH}\cdot{\vd}|^2$ is due to paramagnetism, and is associated with the reduction ($g>0$) of the spin susceptibility for ${\vH}||{\vd}$. By itself the Zeeman energy, $-H_i\delta\chi_{ij}H_j\propto +|{\vH}\cdot{\vd}|^2$, leads to a suppression of $T_c$ for ${\vH}||{\vd}$, but no suppression of $T_c$ for ${\vH}\perp{\vd}$. Machida and Ozaki include this term in order to obtain a tetracritical point in their model. However, there are additional consequences at low temperatures and high fields.  If there were no spin-orbit coupling to lock ${\vd}$ to the crystal lattice, then the superconducting order parameter would nucleate at $H_{c2}$ with ${\vd}\cdot{\vH}=0$, whatever the orientation of ${\vH}$, in order to minimize the Zeeman energy. In the model of Machida and Ozaki the only coupling of ${\vd}$ to the crystal lattice comes through the SBF, which is weak by design, {\ie} $\epsilon M_s^2 \ll g H_{c2}^2$ except for $T\simeq T_{c}$. Thus, at low temperatures and high fields the Zeeman energy dominates the SBF energy.  So for ${\vH}||{\vc}$ the ${\vd}$ vector will nucleate in the basal plane and there will be no Pauli limiting of $H_{c2}^{||}$.  Similarly, for ${\vH}\perp{\vc}$ the ${\vd}$ vector will nucleate in the basal plane, in this case with ${\vd}\perp{\vH}\perp{\vc}$. Thus, there is no paramagnetic limiting at low temperatures. This is an important fact, which appears to conflict with the experimental measurements of the upper critial field.\cite{shi86b,ade90}

\begin{figure}[t]
\begin{overpic}[width=\columnwidth]{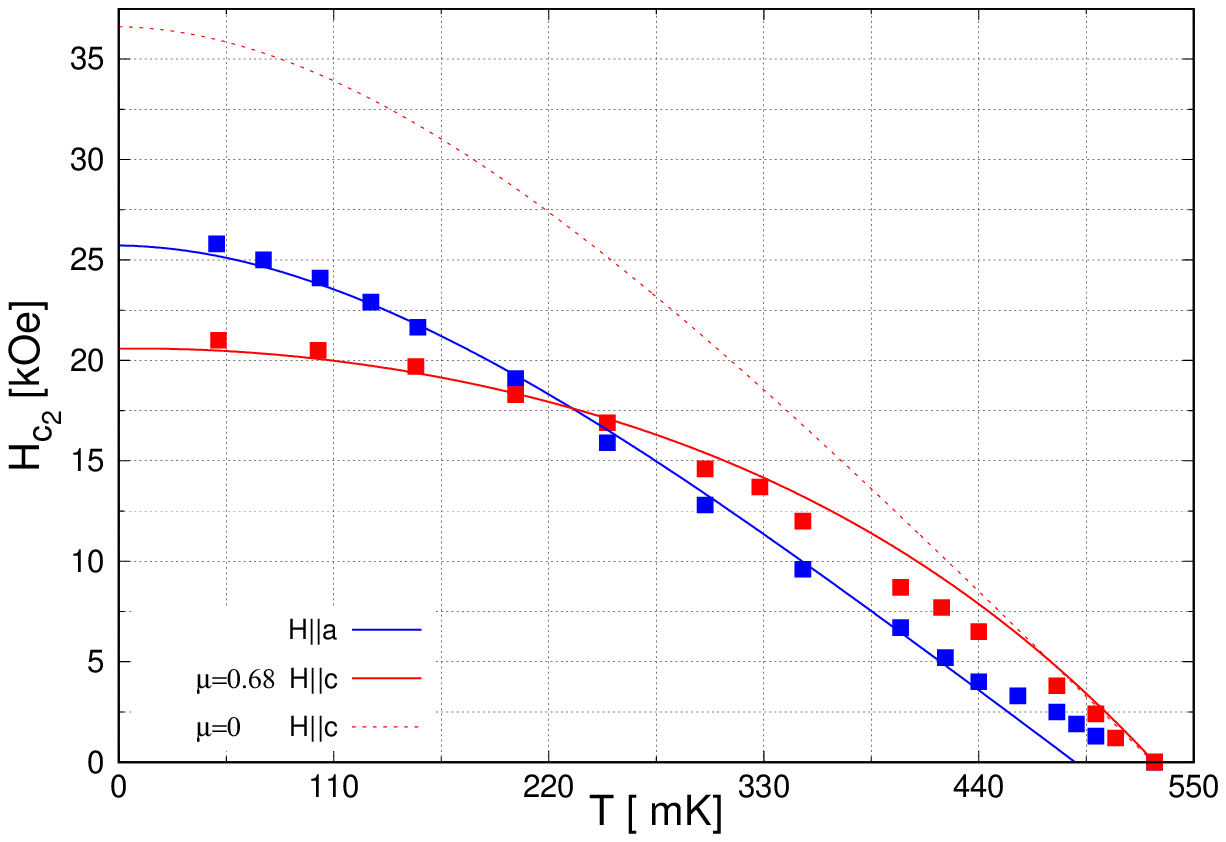}
\put(130,115){\colorbox{white}{\parbox{0.325\columnwidth}{
              \includegraphics[width=0.325\columnwidth]{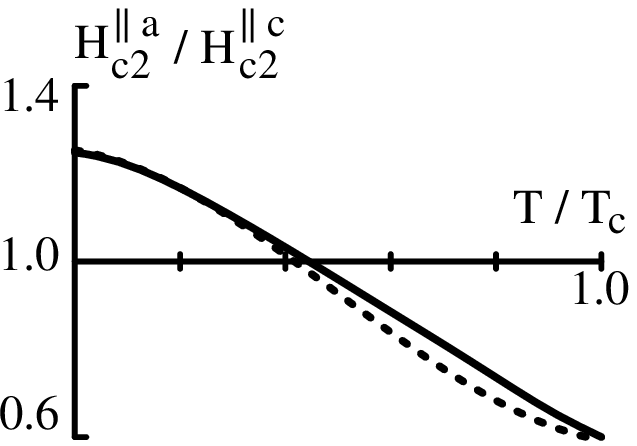}
             }}}
\end{overpic}
\caption{
Anisotropy of the Upper Critical Field. The data is from Ref. (\onlinecite{shi86b})
and the theoretical curves are from Ref. (\onlinecite{cho91}).
}
\label{fig:Hc2_Anisotropy}
\end{figure}

The upper critical field data from Ref. (\onlinecite{shi86b}) for both ${\vH}\perp{\vc}$ and ${\vH}||{\vc}$ is shown in Fig.~\ref{fig:Hc2_Anisotropy}. The unique feature of UPt$_3$ is the cross-over in the anisotropy ratio, $H_{c2}^{\perp}/H_{c2}^{||}$. Estimates of $T_c\,(dH_{c2}/dT)_{T_c}/H_{c2}(0)$ indicate paramagnetic suppression of $H_{c2}$ for ${\vH}||{\vc}$, but {\it no} paramagnetic suppression for ${\vH}\perp{\vc}$. Choi and I obtained a quantitative explanation\cite{cho91} of the upper critical field data\cite{shi86b} within BCS theory and the following inputs:
\begin{enumerate}
\item Uniaxial anisotropy of the Fermi velocity ({\ie} anisotropic coherence lengths, $\xi_{\perp}=\hbar v_f^{\perp}/2\pi T_c > \xi_{||}=\hbar v_f^{||}/2\pi T_c$) determines the anisotropy of $H_{c2}$ near $T_{c}$, while paramagnetic coupling of the quasiparticle pseudospin to the magnetic field, ${\cal H}_{\mbox{\tiny Zeeman}}=-\mu_{\perp}(H_x\sigma_x+H_y\sigma_y)-\mu_{||}\,H_z\sigma_z$, determines the high field, low-temperature anisotropy of $H_{c2}$.
\item The order parameter is \underline{odd-parity} with ${\vd}$ locked to the crystal direction ${\vc}$ by strong spin-orbit coupling. The Zeeman energy is then pair-breaking for ${\vH}||{\vc}$, but {\it not} for ${\vH}\perp{\vc}$. For this latter orientation the field simply shifts the population of Cooper pairs with spin directions 
$\ket{\leftleftarrows}$ and $\ket{\rightrightarrows}$, without any loss of condensation energy; thus, $H_{c2}^{\perp}$ is purely due to diamagnetism.
\item The fit to the data of Shivaram, {\et}\cite{shi86b} yields an effective moment of $\mu_{||}\simeq 0.3 \mu_B$. The calculated temperature dependences of $H_{c2}^{||}$ and $H_{c2}^{\perp}$ yield a cross-over temperature of $\simeq 200\, mK$ in good agreement with the data.
\end{enumerate}

It is important to note that the anisotropy of the upper critical field is not explained by an even-parity order parameter and an anisotropic effective moment tensor even though $\mu_{\perp}\ne\mu_{||}$ in UPt$_3$. The susceptibility anisotropy of the normal state near $T_c$ is $\chi_{||}/\chi_{\perp}=(\mu_{||}/\mu_{\perp})^2\simeq \frac{1}{2}$. Thus, for a conventional singlet gap the anisotropy of the Pauli limit at $T=0$ is estimated to be
\be
\frac{H_P^{||}}{H_P^{\perp}}=\frac{\Delta/\mu_{||}}{\Delta/\mu_{\perp}}
=\frac{\mu_{\perp}}{\mu_{||}}\simeq\sqrt{2}
\,,
\ee
which is the opposite of what is observed. Detailed calculations for the even-parity representations confirm this simple argument; spin-singlet pairing is Pauli limited for all orientations, and the calculated anisotropy of $H_{c2}$\cite{cho91,cho93} is qualitatively inconsistent with the measured anisotropy of $H_{c2}(T)$ at low temperatures and the anisotropy of $\mu_{||}/\mu_{\perp}$.\cite{fri83} Also note that the position of the crossing point of the anisotropy ratio and the magnitude of the anisotropy at $T=0$ are sensitive to impurity scattering. Impurity scattering reduces the anisotropy ratio at $T=0$ and pushes the crossover point to lower temperature; relatively weak disorder ($1/(2\pi\tau T_c = 0.1$) moves the crossover temperature to $T=0$.\cite{cho93}

Our interpretation of the origin of the cross-over in the anisotropy in $H_{c2}$ at low temperature - a spin-triplet state with ${\vd}$ locked to the ${\vc}$ axis by spin-orbit coupling - is in conflict with the model of Ref. (\onlinecite{mac91}). Although Machida and Ozaki\cite{mac91} eliminate the `level repulsion' terms by assuming a 1D orbital representation, the absence of spin-orbit coupling implies that there is no paramagnetic limiting at low temperatures, and thus, no obvious mechanism for generating a cross-over in the anisotropy of $H_{c2}(T)$.
 
\vspace*{-3mm}
\subsection*{Accidentally Degenerate Models}
\vspace*{-3mm}
 
Chen and Garg \cite{che93} recently investigated a GL theory of the phase diagram based on {\it two} primary order parameters belonging to different irreducible representations that are accidentally degenerate, or nearly so (see also Ref. (\onlinecite{joy90,luk91})).  By choosing the two representations appropriately they guarantee that the `level repulsion' terms are absent by symmetry. What is required is that $\eta_a$ and $\eta_b$, corresponding to irreducible representations $a$ and $b$, have different signatures under reflection, or parity. In the simplest case $\eta_a$ and $\eta_b$ are both 1D representations, {\eg} $\eta_a\in A_{2u}$ and $\eta_b\in B_{1u}$. The form of the GL functional is then,

\begin{widetext}
\ber\label{free_energy_ab}
{\cal F} 
&=& 
\int d^3x\,\Big\{
\alpha_a|\eta_a|^2 + \beta_a|\eta_a|^4 +
\alpha_b|\eta_b|^2 + \beta_b|\eta_b|^4 +
b_1|\eta_a|^2|\eta_b|^2 +
b_2(\eta_a\eta_b^* + \eta_a^*\eta_b)^2
\nonumber \\
&&
\qquad\qquad
+\kappa_a|{\vD}_{\perp}\eta_a|^2 + \kappa_b|{\vD}_{\perp}\eta_b|^2 +
\kappa_a'|D_z\eta_a|^2 + \kappa_b'|D_z\eta_b|^2 
\Big\}
\,.
\eer
\end{widetext}

The main features of this model are:
\begin{enumerate}
\item The nearly degenerate double transition occurs because of a near degeneracy of the pairing interaction in two channels unrelated by symmetry.
\item $\kappa_a$ and $\kappa_b$, and similarly for the z-axis derivatives, are independent coefficients in this model.
\item Terms of the form $(D_x\eta_a)(D_y\eta_b)^* + c.c.$ are excluded by symmetry, thus allowing for the possibility of a tetracritical point for all field orientations.
\end{enumerate}

The `accidental degeneracy' model is designed to explain the GL phase diagram, particularly the tetracritical point. However, as Garg and Chen\cite{gar94} point out, without corrections to the GL functional of eq.(\ref{free_energy_ab}) the model is unable to account for a tetracritical point for ${\vH}||{\vc}$. One problem is that the observed tetracritical point occurs at a fairly high field where there is significant curvature in $H_{c2}^{||}(T)$. A related difficulty is that, in contrast to the orientation ${\vH}\perp{\vc}$, there is little or no change in slope of $H_{c2}(T)$ at the tetracritical point.

However, if one assumes that both pairing channels are odd-parity with ${\vd}||{\vc}$, then the paramagnetic correction to the GL functional is
\be
{\cal F}_{\mbox{\tiny para}} = (g_a\,|\eta_a|^2 + g_b\,|\eta_b|^2)\,({\vH}\cdot{\vc})^2
\,,
\ee
corresponding to the suppression of the spin susceptibility for ${\vH}||{\vc}$. One expects both $g_a\,,\,g_b>0$, {\ie} paramagnetism suppresses both order parameters. By themselves the paramagnetic terms would lead to a reduction of $T_{c}$, $T_{{c_a},{c_b}}(H)= T_{{c_a},{c_b}} - (g_{a,b}/\alpha_{a,b}')\,({\vH}\cdot{\vd})^2$, where $\alpha_{a,b}(T)=\alpha_{a,b}'(T-T_{{c_a},{c_b}})$. The key features of the paramagnetic correction are (i) its origin is an odd-parity order parameter with ${\vd}||{\vc}$, as I have argued based on the low-temperature anisotropy of $H_{c2}^{\perp}/H_{c2}^{||}$, (ii) it allows for a tetracritical point for ${\vd}||{\vc}$ with a small change in slope, and (iii) the sharp kink in $H_{c2}^{\perp}(T)$ at the tetracritical point is consistent with the absence of a paramagnetic correction for ${\vH}\perp{\vc}$. The suppression of a kink in $H_{c2}^{||}$ comes about because the paramagnetic suppression of $H_{c2}$ is dominant on the low temperature, high-field side of the tetracritical point. To leading order in $g$ (I assume $g_a =g_b=g$) the ratio of the slopes of $H_{c2}^{||}(T)$ above and below the tetracritcal point are
\be
\frac{(-dH_{c2}/dT)^{<}_{K}}{(-dH_{c2}/dT)^{>}_{K}}
=
\frac{\kappa_a}{\kappa_b}
\left(1 - \frac{gH_{K}\phi_0}{\pi}\,
\frac{\kappa_a - \kappa_b}{\kappa_a\kappa_b}\right)
\,,
\ee
where phase $b$ is the high-field, low-temperature phase, and $H_{K}$ is the field at the tetracritical point. In the absence of of paramagnetism the ratio of slopes is given by $\kappa_a/\kappa_b$.  As expected paramagnetism smooths the kink out. Paramagnetism also moves the tetracritcal point to lower temperatures, $T_{K}=T_{K0}[1-gH_{K}\phi_0/2\pi(1/(\kappa_a+\kappa_b))]$. With the paramagnetic correction added to the model of Ref. (\onlinecite{che93}) it is possible to account for the slopes of $H_{c2}(T)$ and the positions of the tetracritical point of Ref.(\onlinecite{ade90}). My analysis of the phase diagram within this model, which allows for a tetracritical point for $H_{c2}^{||}$, with a very small slope discontinuity as a result of paramagnetic suppression, gives $|(\kappa_a - \kappa_b)/2\kappa_a|$ and $|(\kappa_a' - \kappa_b')/2\kappa_a'|\simeq 0.1-0.2$.\footnote{There is uncertainty in these coefficients partly associated with the assumed values for the g-factors and partly because the higher order gradient corrections to the GL free energy are formally the same order in $\sqrt{1-T/T_c}$ as the paramagnetic corrections. Also, the determination of the in-plane stiffness coefficients ($\kappa_a$ and $\kappa_b$) from $H_{c2}^{||}(T)$, indirectly affects the determination of $\kappa_a'$ and $\kappa_b'$ from $H_{c2}^{\perp}(T)$. However, if I assume that the differences, $(\kappa_a-\kappa_b)/2\kappa_a$ and $(\kappa_a'-\kappa_b')/2\kappa_a'$ are roughly the same, then I obtain a ratio of $\sim 0.15$ and can account for the tetracritical point for both field orientations. Garg and Chen\cite{gar94} assume $\kappa_a=\kappa_b$, in which case all of the slope discontiunity for $H_{c2}^{\perp}$ must be accounted for by the difference in the ${\vc}$-axis stiffness coefficients; thus, $(\kappa_a'-\kappa_b')/2\kappa_a'\simeq 0.3$. In addition, in order to account for the tetracritical point with $\kappa_a=\kappa_b$, one has to take $g_a\gg g_b$ for the two odd-parity amplitudes.}

While there is sufficient structure in this model to account for the features of the H-T phase diagram, the accidental degeneracy model does not account for the correlation between superconductivity and AFM that has been found in pressure studies.  Another potential difficulty is that several experiments report power law temperature dependences for transport coefficients at low temperature ($T\ll T_c$) that are consistent with a line of nodes in the basal plane (see below). None of the accidental degeneracy models based on two 1D representations exhibit line nodes of the excitation gap parallel to the basal plane in the clean limit (see Table \ref{tab_OPs}). For example, the odd-parity model $(A_{2u}+iB_{1u})$ with ${\vd}||{\vc}$ has a gap with six line nodes  {\it perpendicular} to the basal plane.

\vspace*{-3mm}
\subsection*{The E$_{2u}$ model}
\vspace*{-3mm}

Although a model based on two primary order parameters is capable of explaining the existing experimental data for the phase diagram, when one considers the BCS relation between the effective interaction and the transition temperature, $T_{c}=\omega_c\,\exp\{-\frac{1}{V}\}$, an accidental degeneracy of two pairing channels at the level of a few percent seems implausible. However, a primary order parameter belonging to a single higher dimensional representation, which is coupled to a weak symmetry breaking field, provides a natural explanation for two superconducting phases with nearly degenerate transition temperatures.  Here I argue that the SBF explanation for the double transition based on a 2D orbital representation is not ruled out by the `topological isotropy of the tetracritical point'. In addition, I show how the apparent tetracritical point can arise from the SBF in this theory.

Although the GL theories are formally the same for any of the 2D orbital representations, the predictions for the GL material parameters differ substantially depending on the symmetry of the Fermi surface and the Cooper pair basis functions. For example, the interpretation of the $H_{c2}$ and susceptibility anisotropy in terms of anisotropic Pauli limiting requires an odd-parity, spin-triplet representation with the ${\vd}$-vector parallel to the ${\vc}$ direction. This limits us to either the E$_{2u}$ or E$_{1u}$ basis functions among the four possible
2D representations.

\vspace*{-12mm}
\subsubsection*{BCS predictions and the `level repulsion' terms}
\vspace*{-3mm}

There are other important predictions from the weak-coupling BCS theory for the 2D representations.  For any of the four 2D representations, the fourth-order free energy coefficients have the ratio, $\frac{\beta_2}{\beta_1}=\frac{1}{2}$. This result was reportedfor the $E_{1g}$ representation based on a clean-limit calculation and a spherical Fermi surface.\cite{sch89} The more general result is that $\beta_2/\beta_1 =\frac{1}{2}$ is also insensitive to hexagonal anisotropy of the Fermi surface and basis functions, and to non-magnetic, s-wave impurity scattering. Although impurity scattering is pair-breaking for any of the 2D representations, the impurity renormalization of the $\beta$'s drops out of the ratio $\beta_2/\beta_1$ for s-wave impurity scattering. This result ensures that the coupling of the SBF to the superconducting order parameter will produce a double transition in zero field for any of the 2D orbital representations.

Significant differences between the 2D models appear when we consider the gradient terms in the GL functional, or equivalently the GL differential equations, calculated from BCS theory. The gap equation is given by the mean-field BCS equation,

\ber
\Delta_{\alpha\beta}({\vk}_f,{\vx})
&=&
\int\,d^2{\vk}_f'\,n({\vk}_f')\,
V_{\alpha\gamma;\beta\rho}({\vk}_f,{\vk}_f')\,
\nonumber\\
&\times&
T\sum_{\epsilon_n}^{|\epsilon_n|<\omega_c}\,
f_{\gamma\rho}({\vk}_f',{\vx};\epsilon_n)
\,,
\eer
where $V_{\alpha\gamma;\beta\rho}({\vk}_f,{\vk}_f')$ is the pairing interaction, $n({\vk}_f)$ is the angle-resolved density of states on the Fermi surface and $f({\vk}_f,{\vx};\epsilon_n)$ is the quasiclassical pair amplitude at Matsubara frequencies and momenta on the Fermi surface. In the weak-coupling limit the interaction is cutoff at a frequency $T_c\ll\omega_c\ll E_f$, and both the cutoff and the pairing interaction are eliminated with the linearized, homogeneous gap equation in favor of $T_c$. In order to calculate the leading order gradient terms in the GL equation we need only examine the linearized gap equation. To linear order in $\Delta$ the pair amplitude can be calculated straight-forwardly from the quasiclassical transport equations\cite{rai94}; in the clean limit $f$ satisfies the differential equation (see {\eg} Ref. (\onlinecite{cho89}),
\ber
\left\{\frac{|\epsilon_n|}{\pi}
+
\frac{{\rm sgn}(\epsilon_n)}{2\pi}\,{\vv}_f\cdot{\vD}\right\}\,f=\,\Delta
\,.
\eer
Near $T_c$ the estimates $|\epsilon_n|\sim T_c$, $|{\vv}_f\cdot{\vD}|\sim T_c\sqrt{1-T/T_c}$ apply, so that to leading order in gradients the linearized equation for the odd-parity gap function becomes,
\ber
{\vDelta}({\vk}_f,{\vx})
&=&
\int\,d^2{\vk}_f'\,n({\vk}_f')\,{\bf V}({\vk}_f,{\vk}_f')\cdot
\\
&\times&
\left\{{\cal K}(T) + \frac{7\zeta(3)}{16\pi^2T_c^2}\,({\vv}_f'\cdot{\vD})^2\right\}\,
{\vDelta}({\vk}_f',{\vx})
\,,
\nonumber
\eer
where ${\cal K}(T)=\ln(1.13\omega_c/T)$ and ${\bf V}({\vk}_f,{\vk}_f')$ is the pairing interaction in the odd-parity, spin-triplet channel. The same equation holds for the even-parity channel with the appropriate substitutions for the gap function and pairing interaction. This equation is used to generate the coefficients of the gradient terms in the GL equations. For the even-parity, or odd-parity with ${\vd}||{\vc}$, 2D models I obtain in the clean limit\cite{sau94}
\ber\label{kappas}
\kappa_1= & \kappa_{0}\,
\left<{\cal Y}_1({\vk}_f)\,v_{fy}\,v_{fy}\,{\cal Y}_1({\vk}_f)\right>
\nonumber \\
\kappa_2= & \kappa_3=
\kappa_{0}\,
\left<{\cal Y}_1({\vk}_f)\,v_{fx}\,v_{fy}\,{\cal Y}_2({\vk}_f)\right>
\nonumber \\
\kappa_4= & \kappa_{0}\,
\left<{\cal Y}_1({\vk}_f)\,v_{fz}\,v_{fz}\,{\cal Y}_1({\vk}_f)\right>
\,,
\eer
where ${\cal Y}_i({\vk}_f)$ are the basis functions,
$\kappa_{0}=\frac{7\zeta(3)}{16\pi^2 T_c^2}\,N_f$, and $N_f$ is the
density of states at the Fermi level. There are important differences
between the the E$_{1}$ and E$_{2}$ representations when we evaluate
these averages for the in-plane stiffness coefficients. In the clean
limit, using the basis functions in Table \ref{tab_basis} and a Fermi
surface with weak hexagonal anisotropy,
the E$_{1}$ model gives
\be
\kappa_2=\kappa_3\simeq\kappa_1 \qquad\qquad\qquad({\rm E}_{1})
\,,
\ee
while for the E$_{2}$ model I obtain\footnote{This statement can be made quantitative with a specific bandstructure. Consider the model bandstructure, $\varepsilon_{{\vk}}=\frac{1}{2m_{\perp}}(k_x^2+k_y^2) + \frac{1}{2m_{||}}\,k_z^2 + \alpha\,Re(k_x+ik_y)^6$, which represents the lowest-order hexagonal correction to a cylindrically symmetric dispersion relation. The Fermi surface is defined by $\varepsilon_{{\vk}_f}=\mu$; it is convenient to parametrize the Fermi wavevector in terms of an angle-dependent magnitude and the direction cosines, $(\hat{k}_x,\hat{k}_y,\hat{k}_z)$.  To leading order in $\alpha$ the in-plane Fermi velocity becomes $v_x=v_{\perp}(1+ 3\varpi[\hat{k}_x^4-10\hat{k}_x^2\hat{k}_y^2+5\hat{k}_y^4])\hat{k}_x$ and $v_y=v_{\perp}(1- 3\varpi[\hat{k}_y^4-10\hat{k}_y^2\hat{k}_x^2+5\hat{k}_x^4])\hat{k}_y$, with $\varpi=\alpha k_0^6/\mu$ and $k_0=(2m_{\perp}\mu)^{\frac{1}{2}}$.  The resulting stiffness coefficients become $\kappa_2=\kappa_3=-0.56\varpi\,\kappa_0\,v_{\perp}^2$ implying $|\kappa_{2,3}|\sim |\varpi|\kappa_1\ll\kappa_1$. A detailed analysis of these gradient coefficients could be carried out for the multi-sheet Fermi surface of UPt$_3$.  However, without a microscopic model for the pairing mechanism, or other specific information regarding the detailed anisotropy of the order parameter, it is not clear what weight to attach to the pairing amplitude on different sheets of the Fermi surface.  The indirect information we have is (i) the in-plane isotropy of the upper critical field at low temperatures, which suggests that little if any hexagonal anisotropy contributes to Fermi surface averages in eqs. (\ref{kappas}), and (ii) the extrapolation of the heat capacity ratio, $\lim_{T=0}C/T\rightarrow\gamma_s$.  The latter could be interpreted as a gapless region of the Fermi surface, possibly due to negligible pairing amplitude on one or more sheets of the Fermi surface.\label{foot:E2u}}
\be
\kappa_2=\kappa_3\ll\kappa_1\sim N_f\left(\frac{v_f^{\perp}}{\pi T_c}\right)^2\quad({\rm E}_{2})
\,.
\ee
In fact, the three in-plane coefficients are identical for E$_{1}$ in the limit where the in-plane hexagonal anisotropy of the Fermi surface vanishes. In contrast, the coefficients $\kappa_2$ and $\kappa_3$ for the E$_{2}$ model both vanish when the hexagonal anisotropy of the Fermi surface is neglected. This latter result follows directly from the approximation of a cylindrically symmetric Fermi surface and Fermi velocity, ${\vv}_f=v_f^{\perp}(\hat{k}_x\,{\vx}+\hat{k}_y\,{\vy})+v_f^{||}\hat{k}_z{\vz}$, and the higher angular momentum components of the E$_{2}$ basis functions,
\be
\hspace*{-2mm}
\kappa_2(E_{2u})
\propto\langle\hat{k}_z(\hat{k}_x^2\ns-\ns\hat{k}_y^2)v_{fx} v_{fy}
(2\hat{k}_x \hat{k}_y)\hat{k}_z\rangle \equiv 0
\,.
\ee
This is a crucial point; if there is weak hexagonal anisotropy then $\kappa_{23}\ll\kappa_1$ only for E$_{2}$.\cite{Note11} The conclusion is that there is a natural explanation for the absence (or at least the smallness) of the `level repulsion' terms in the orbital 2D model, but we are required to select the E$_{2}$ representation and have weak hexagonal anisotropy of the Fermi velocity in the basal plane. There is a support for this latter assumption; if the hexagonal anisotropy of ${\vv}_f$ were significant it should be observable at low temperature as an in-plane anisotropy of $H_{c2}^{\perp}(T)$. The angular dependence of $H_{c2}^{\perp}$ at low temperatures was investigated, but no in-plane anisotropy was observed.\cite{shi86b}

In order to account for the discontinuities in the slopes of the transition lines near the tetracritical point I need an additional ingredient in the GL theory for the E$_{2u}$ model that is not present in the theory of Hess, {\it et al.}\cite{hes89} For E$_{2u}$ with $\kappa_{23}=0$ the gradient energy reduces to
\begin{widetext}
\be
{\cal F}_{\mbox{\tiny grad}}=\int\ns d^3x 
\Big\{
\kappa_1\left(|{\vD}_{\perp}\eta_1|^2 + |{\vD}_{\perp}\eta_2|^2\right) 
+
\kappa_4\left(|D_z\eta_1|^2 + |D_z\eta_2|^2\right)
\Big\}
\,.
\ee
\end{widetext}
Because both order parameter components appear with the same coefficients there is no crossing of different $H_{c2}(T)$ curves corresponding to different eigenfunctions, and therefore no apparent tetracritical point. However, the analysis of the slopes of the transition lines near the tetracritical point (see above) suggests that the difference in the gradient energies associated with the two components of the order parameter are finite, but small, {\ie} $|\Delta\kappa/2\kappa_1|\lesssim 0.2$. This suggests that the SBF may be responsible for a splitting in the gradient coefficients as well as the transition temperature.

In the model of Hess, {\et}\cite{hes89} the coupling to the SBF was included through second order in both the superconducting order parameter, ${\veta}$, and the AFM order parameter, ${\vM}_s$, but only for the homogeneous terms in the free energy. The motivation in the original paper was to provide a mechanism for the double phase transition in zero field. The second-order contribution of the SBF to the gradient energy was not included. In retrospect, these terms are as essential for describing a double transition as a function of field, as the homogeneous term is for the double transition in zero field. The relevant invariants can be generated by the simple algorithm,
\be\label{algorithm}
\eta_i \rightarrow (\delta_{ij} + \frac{1}{2} \epsilon M_i M_j)\eta_j
\,,
\ee
in eq.(\ref{free_energy}). To second-order in ${M}_s$ the homogeneous
coupling to the SBF is generated,
\be
\alpha (|\eta_1|^2+|\eta_2|^2)\rightarrow
\alpha(1 + \epsilon M_s^2)\,|\eta_1|^2 +
\alpha(1 - \epsilon M_s^2)\,|\eta_2|^2
\,,
\ee
which accounts for the double transition in zero field.  The SBF coupling to the order parameter also contributes at second-order to the gradient energy,
\be
\kappa_1(|{\vD}_{\perp}\eta_1|^2 + |{\vD}_{\perp}\eta_2|^2)
\rightarrow
(\kappa_1^{+}|{\vD}_{\perp}\eta_1|^2 + 
\kappa_1^{-}|{\vD}_{\perp}\eta_2|^2)
\,,
\ee
where
\be
\kappa_1^{\pm} = \kappa_1 (1 \pm \epsilon_{\perp} M_s^2)
\,,
\ee
and similarly for the ${\vc}$-axis gradients,
\be
\kappa_4^{\pm} = \kappa_4(1 \pm \epsilon_{||} M_s^2)
\,.
\ee
It should be noted that the the replacement in eq.(\ref{algorithm}) is an expedient algorithm for generating the couplings to the SBF.  Symmetry analysis yields the same invariants, in addition to other corrections of order $M_s^2$ which I ignore here.\cite{sau94b} The coupling coefficients, $\epsilon$, $\epsilon_{\perp}$, $\epsilon_{||}$, for the homogeneous term, the in-plane gradient energies and the ${\vc}$-axis gradient energies are not identical. In the absence of a microscopic calculation of these coupling parameters dimensional analysis implies that they are formally the same order of magnitude, in which case we conclude that the splittings in the gradient coefficients are relatively small,
\be
\left|\frac{\kappa_{1,4}^{+}-\kappa_{1,4}^{-}}{2\kappa_{1,4}}\right|=
\left|\epsilon_{\perp,||}\,M_s^2\right|\sim\left|\frac{\Delta T_c}{T_c}\right|
\,,
\ee
which is consistent with the analysis of the tetracritical point.\cite{Note10} Thus, within the E$_{2u}$ model the SBF is essential for producing an apparent tetracritical point, and at a semi-quantitative level, can account for the magnitudes of the slopes near the tetracritical point.

\vspace*{-3mm}
\subsubsection*{Nodes in the gap}
\vspace*{-3mm}

In addition to providing a reasonable description of the phase diagram, the E$_{2u}$ model also has the geometry for the nodes of the excitation gap that accounts qualitatively for the temperature dependences of the acoustic attenuation and penetration depth at low temperatures. The existence of a line node in the basal plane has been argued by several authors,\cite{sch86,bro90}, and it has been assumed to favor the even-parity E$_{1g}$ order parameter of the form $\Delta_{E_{1g}}\sim k_z(k_x + i k_y)$.\cite{bro90} There is not yet consistency between the predicted transport properties, the assumed nodal structure of the excitation gap and the experimental results for several different transport measurements.\cite{vor92} However, the presence of a line node in the basal plane appears to be reasonably well established from transverse ultrasonic absorption measurements.\cite{shi86a} As Norman points out\cite{nor92} the E$_{2u}$ order parameter for the low-temperature phase,
\be\label{E2u_0}
{\vDelta}\sim{\vc}\,\, \hat{k}_z\,(\hat{k}_x + i \hat{k}_y)^2
\,,
\ee
has a line of nodes in the basal plane as well as point nodes along the ${\vc}$-axis.\footnote{The topology of the low-temperature excitation gap is not changed by the perturbative corrections from the SBF.}
However, the interpretation of the temperature dependences directly in terms of the order parameter is complicated by material effects, particularly impurity scattering.\cite{pet86,hir86,sch86} Also note that even though the topology of the nodes for E$_{1g}$ and E$_{2u}$ are the same there is a difference in the excitation spectrum near the point nodes in the two cases; the spectrum opens linearly with polar angle for the E$_{1g}$ state and quadratically for the E$_{2u}$ state, and will give rise to a corresponding difference in the anlgle-resolved density of states near the polar nodes. Thus, a thorough examination of the E$_{2u}$ model and the experimental data on the low-temperature superconducting properties is required before any stronger conclusions can be drawn about whether or not the E$_{2u}$ model can account for the low-temperature transport properties.

On a different, but related aspect, of `nodes in the gap' the interpretation of the low-temperature transport and thermodynamic data in terms of a line of nodes in the excitation gap, combined with the group-theoretical analysis of several authors,\cite{vol84,blo85} has been used to argue in favor of an even-parity order parameter in UPt$_3$.\cite{vol85} However, the realization of an odd-parity order parameter with a line of nodes in the gap, even with strong spin-orbit coupling, does not violate any rigorous group-theoretical result. The result of Volovik and Gorkov\cite{vol84} and  Blount\cite{blo85} is that symmetry does not enforce a line of nodes for odd-parity gaps.  However, if the pairing interaction (because of spin-orbit coupling) selects an `easy axis' for the ${\vd}$ vector, {\ie} ${\vd}||{\vc}$ as I have argued based on the upper critical field data (Norman also argues for ${\vd}||{\vc}$ based on his spin-fluctuation model for the pairing interaction\cite{nor91}), then the odd-parity, E$_{2u}$ basis functions necessarily have a line of nodes in the basal plane.\cite{yip93c}

\vspace*{-5mm}
\subsection*{Tests of the Order Parameter}
\vspace*{-3mm}

The E$_{2u}$ model, as well as other models for UPt$_3$, exhibit a number of symmetries and broken symmetries that can, in principle, be used to uniquely identify, or eliminate, any one model as the order parameter for the phases of UPt$_3$. I include a discussion of some further tests of the order parameter, some of which are `crucial tests' in the sense that directly test for a broken symmetry or a residual symmetry of the order parameter.

\vspace*{-3mm}
\subsubsection*{Meissner Effects}
\vspace*{-3mm}

An important feature of the quasiparticle excitation spectrum in most unconventional superconductors is that the gap vanishes along lines or at points on the Fermi surface. These gapless regions imply low-energy excitations, at all temperatures, which give rise to power law temperature dependences for the penetration depth for $T\ll T_c$. The observation of non-activated behavior for $\lambda(T)$ at $T\ll T_c$ is often interpreted as evidence for nodes in excitation gap.\cite{gro86,bro90,sig92}

The nodes in the excitation spectrum also lead to anomalies in the velocity-dependence of supercurrent,\cite{vol81b,muz83,yip92a} which should be observable at very low temperatures in the field-dependence of the penetration depth.\cite{yip92a} The importance of the nonlinear Meissner effect is that it is particularly sensitive to the {\it positions} of the nodes in ${\vk}$-space, and could in principle be used to distinguish between the gaps in Table \ref{tab_OPs}. The origin of this field dependence is obtained by considering a clean superconductor with ellipsoidal Fermi surface and an E$_{2u}$ order parameter given by eq. (\ref{E2u_0}), which has a line of nodes in the basal plane and point nodes on at the upper and lower positions of the Fermi surface.

In the presence of the condensate flow field, ${\vv}_s ={1\over 2}({\vpartial}\chi+{2e\over \hbar c}{\vA})$, the energy of a quasiparticle at the position ${\vk}_f$ on the Fermi surface is given by $E+{\vv}_f\cdot{\vv}_s$, where $E= (\epsilon^2+|\Delta({\vk}_f)|^2)^{1/2}$ and $\epsilon$ is the quasiparticle energy in the normal state. The equilibrium distribution of quasiparticles is therefore $f(E+{\vv}_f\cdot{\vv}_s)$. Consider the geometry where ${\vv}_s$ is directed in the basal plane.  The important point is this: at $T=0$, for any non-zero ${\vv}_s$ there is a wedge of {\it occupied} states near the node opposite to the flow velocity. Thus, the supercurrent is reduced even at $T=0$ from the ideal value for pure condensate flow by a backflow correction of order $({v_f v_s\over 2\Delta_o})$. The net supercurrent is easily calculated from the phase space of occupied states to be,
\begin{equation}
{\vj}_s=-{e\over 2}N_f (v^{\perp}_f)^2\,{\vv}_s\
\left\{1-{|{\vv}_s|\over 2\Delta_o/v^{\perp}_f}\right\}\ ,
\label{cur_line}
\end{equation}
for ${\vv}_s$ in the basal plane. Here $\Delta_o$ is related to the rate at which the gap opens up at the nodes; for simplicity I assume the gap function $|\Delta({\vk}_f)| = \Delta_o |k_z|(\hat k_x^2 + \hat k_y^2)$. Note that the velocity dependence of the effective superfluid density, $\rho_s=\rho_o \{1-{|{\vv}_s|\over 2\Delta_o/v_f}\}$, is linear and non-analytic, in contrast to the quadratic behavior expected for backflow from thermally excited quasiparticles.

The nonlinear current-velocity relation, in the clean limit, reflects the position and dimensionality of the nodes in the excitation gap, and implies a similar behavior for the field dependence of the penetration length,\cite{yip92a}
\begin{equation}
\left({1\over\lambda_{\perp}}\right)^{\mbox{\tiny eff}} 
=
{1\over\lambda_{\perp}}\left\{1-{H\over H_o}\right\}
\,,\,{\vn}||{\vc}\perp{\vH}\,,
\end{equation} 
where $H_o\sim\phi_o/(\lambda\xi)\sim H_c$ and ${\vn}$ is the normal to the surface.  Finite temperature effects produce a low-field cross-over for current flow in the basal plane as a result of redistribution of thermal quasiparticles in the flow field. Below this cross-over $\lambda_{\mbox{\tiny eff}}$ becomes quadratic in $H$ for all orientations. The cross-over field is estimated by equating the excitation energy of a thermal quasiparticle with the shift in the quasiparticle energy associated with the superflow, $\pi T\simeq v_fv_s\simeq \Delta_o(H_x/H_o)$; {\ie} $H_x\simeq H_o(T/T_c)$.

The conditions for observing the linear field dependence associated with the zero-temperature anomaly in the Meissner current depend on several factors; (i) minimizing thermal quasiparticle backflow, (ii) reducing impurity scattering, which modifies the DOS near the nodes, and (iii) suppressing vortex nucleation. The latter effect requires fields below the vortex nucleation field, $H_{c1}\simeq 100\,G$, which should be compared with the field scale $H_0\simeq \frac{\phi_0}{\pi\xi^2} (\frac{\xi}{\lambda_{\perp}})\simeq\,1\,kG$. At $T\simeq 5\,mK$ ($T/T_c\simeq 0.01$) thermal quasiparticles are negligible except for very low fields. The cross-over field, below which the thermal backflow dominates the non-thermal quasiparticle backflow current, is approximately $H_x\simeq(T/T_c)H_0\simeq 0.1\,H_{c1}$. Thus, at this temperature there is a sizeable window of fields below $H_{c1}$ which is dominated by the non-thermal backflow current. Furthermore, the resolution of the non-thermal current in the penetration depth should be observable; the change in penetration depth over the field range from zero to $H_{c1}$ is of order $\frac{\delta\lambda}{\lambda}=\frac{H_{c1}}{H_0}\simeq10\,\%$. Observation of an isotropic linear field dependence of the low-temperature in-plane penetration depth would provide strong evidence for a line of nodes in the basal plane, and argue against those models with an array of point nodes or line nodes perpendicular to the basal plane. I summarize in Table \ref{tab_OPs} the basic structure of the excitation gap in the low-temperature phases of the models discussed here, in addition to their residual symmetries and degeneracies which I discuss in the following sections.

\begin{widetext}
\begin{table}
\caption{Low-temperature phases of several models}
\hspace*{-5mm}\begin{minipage}{1.05\textwidth}
\label{tab_OPs}
\begin{tabular}{clclc}
$^a$Pairing Channel
& 
Order Parameter
& 
$^b$Symmetry (H)
& 
\hspace{5mm}$^c$Nodes 
& 
$^d$SQUID phase shift
\\ 
\hline
E$_{1u}$ 
& 
$(k_x + i k_y)\equiv k_+$ 
& 
$D_{6}[E]$ 
&
$\theta=0\,,\pi$ (p) 
& 
$\frac{2\pi}{3}$\\ 

E$_{2u}$ 
& 
$k_z\,(k_+)^{2}$ 
& 
$D_{6}[C_2]$ 
&
$\theta=0\,,\pi$ (p); $\theta=\frac{\pi}{2}$ (l) 
& $\frac{4\pi}{3}$ \\

$A_{1u}\oplus B_{1u}$ 
& 
$[A\,k_z+iB\,Im(k_+)^{3}]$ 
& 
$D_{6}[D_3']$ 
&
$\theta=\pi/2$ \& $\phi_n=n\frac{\pi}{3}$ (p)
& 0 \\

$A_{1u}\oplus B_{2u}$ 
& 
$[A\,k_z+iB\,Re(k_x + i k_y)^{3}]$ & $D_{6}[D_3]$ 
&
$\theta=\pi/2$ \& $\phi_n=(2n+1)\frac{\pi}{6}$ (p)& 0\\

$A_{2u}\oplus B_{1u}$ 
& 
$[A\,k_z\,Im(k_+)^{6} + iB\,Im(k_+)^{3}]$ 
& 
$D_{6}[C_3]$ 
&
$\theta=0,\pi$ (p)\,,$\phi_n=n\frac{\pi}{3}$ (l)
& 
0 \\

$A_{2u}\oplus B_{2u}$ 
& 
$[A\,k_z\,Im(k_+)^{6}+iB\,Re(k_+)^{3}]$ 
& 
$D_{6}[C_3]'$ 
&
$\theta=0,\pi$ (p)\,,$\phi_n=(2n+1)\frac{\pi}{6}$ (l)
& 
0 \\ 

${\cal D}_{1}\otimes A_{1u}$
& 
$k_z$
& 
$D_6[C_6]$ 
&
$\theta=\frac{\pi}{2}$ (l) 
& 
0 \\

${\cal D}_{1}\otimes A_{2u}$
& 
$k_z\,Im(k_+)^{6}$
& 
$D_6[C_6]'$ 
&
$\theta=\frac{\pi}{2}$ (l)\,, $\theta=0\,,\pi$ (p) \,,$\phi_n=n\frac{\pi}{6}$ (l)
& 
0 \\

${\cal D}_{1}\otimes B_{1u}$
& 
$Im(k_+)^{3}$ 
& 
$D_6[D_3']$ 
&
$\theta=0,\pi$ (p)\,,$\phi_n=n\frac{\pi}{3}$ (l)
& 
0 \\

${\cal D}_{1}\otimes B_{2u}$
& 
$Re(k_+)^{3}$
& 
$D_6[D_3]$ 
&
$\theta=0,\pi$ (p)\,,$\phi_n=(2n+1)\frac{\pi}{6}$ (l)
& 0
\\
\hline
\end{tabular}
\footnotetext[1]{The first six entries are based on strong spin-orbit coupling with $\vd\parallel\hat\vc$ and the symmetry group $[D_{6h}]_{\mbox{\tiny spin-orbit}}\times{\cal T}\times U(1)$. The last four entries are based on no spin-orbit coupling and the group $SO(3)_{\mbox{\tiny spin}}\times [D_{6h}]_{\mbox{\tiny orbit}}\times{\cal T}\times U(1)$. ${\cal D}_{1}$ refers to the $S=1$ representation of the rotation group.} 
\footnotetext[2]{The notation for the residual symmetry group follows Ref.(\onlinecite{vol85}), {\it e.g.} $D_{6}[E]$ refers to the combined group composed of the elements of $D_{6}$ with properly chosen elements from $U(1)$ and ${\cal T}$. $[E]$ implies that without elements from these groups rotational symmetry would be completely broken. For all the cases listed the orbital group listed is combined with the two element group, $C_i[E]=\{E,e^{i\pi}C_i\}$, and for the cases without spin orbit coupling the full residual symmetry group also includes the symmetry operations of the ${\vd}$-vector, $\sim{\vx}+i{\vy}$, {\it i.e.} the spin rotation group, $D_{\infty}[E]_{\mbox{\tiny spin}}$.} 
\footnotetext[3]{(p) denotes point nodes, and (l) denotes line nodes.} 
\footnotetext[4]{The SQUID phase shifts correspond to the offset in the maximum Josephson supercurrent, $I_{\mbox{\tiny max}}[\Phi/\Phi_0]$ for the geometry of Fig.~\ref{fig:SQUID}}
\end{minipage}
\end{table}
\end{widetext}

\vspace*{-3mm}
\subsubsection*{Josephson Effects: Gauge-Rotation Symmetry}
\vspace*{-3mm}

The a.c. Josephson effect is arguably the most striking manifestation of broken gauge symmetry in superconductors. In an unconventional order parameter qualitative changes in the current-phase relation can occur which reflect the residual symmetry group of the order parameter.\cite{sau85,ges85,mil88} The three classes of models listed in Table \ref{tab_OPs} are distinguished by their residual symmetry groups.

\begin{figure}
\includegraphics[width=\columnwidth]{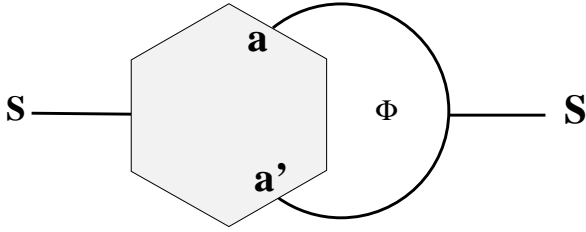}
\caption{
SQUID geometry for UPt$_3$/S junctions.
}
\label{fig:SQUID}
\end{figure}

Consider the E$_{2u}$ and E$_{1u}$ order parameters with ${\vd}||{\vc}$. Both states break gauge symmetry and rotational symmetry, but preserve six-fold {\it gauge-rotation symmetry}; thus, the residual symmetry group is composed of all the six-fold rotations properly combined with gauge transformations. The Josephson current-phase relation is sensitive to the basic gauge transformation associated with the residual symmetry group.\cite{ges85} Under a $60^{\circ}$ rotation about the $\vc$ axis the $E_{2u}$ order parameter acquires a phase
\be
{\vDelta}(E_{2u})\sim {\vc}\,\,\hat{k}_z(\hat{k}_x+i\hat{k}_y)^2\,
{\buildrel {\pi/3}\over \longrightarrow}\,\,e^{i2\pi/3}{\vDelta}(E_{2u})
\,,
\ee
that is twice that acquired by the $E_{1u}$ order parameter for the same rotation,
\be
{\vDelta}(E_{1u})\sim {\vc}\,\,(\hat{k}_x+i\hat{k}_y)\,
{\buildrel {\pi/3}\over \longrightarrow}\,\,e^{i\pi/3}{\vDelta}(E_{2u})
\,.
\ee
Now consider the geometry of Fig.~\ref{fig:SQUID} with two junctions between $UPt_3$ (assuming either an E$_{1u}$ or E$_{2u}$ order parameter) and a conventional s-wave superconductor on two different faces, $a$ and $a'$, of a hexagonal crystal. The junctions are related by a $120^{\circ}$ rotation, but are otherwise identical. Under a $120^{\circ}$ rotation the order parameter undergoes a phase change. Equivalently, the $120^{\circ}$ rotation followed by a gauge transformation of $\phi_u\rightarrow\phi_u-2\mu\pi/3$ (with $\mu=1$ and $2$ for the E$_{1u}$ or E$_{2u}$ representations, respectively) is a symmetry operation. Thus, for the supercurrents at
$a$ and $a'$,
\begin{equation}
I_a(\phi_u-\phi_s)=I_{a'}(\phi_u-\phi_s+{2\mu \pi \over 3})
\label{squid}
\,,
\end{equation}
where $\phi_s$ is the phase of the s-wave order parameter. This symmetry has an interesting experimental consequence. Consider the SQUID constructed from these junctions (Fig.~\ref{fig:SQUID}). Equation (\ref{squid}) implies that the maximum critical current for the SQUID occurs for an external flux $\Phi = (n + {\mu \over 3}) \Phi_o$, where $n$ is an integer and $\Phi_o=\frac{hc}{2 e}$ is the flux quantum. This phase shift of the interference pattern is a signature of residual gauge-rotation symmetry and allows us to differentiate between E$_{1u}$, E$_{2u}$ and the other order parameters discussed as models of UPt$_3$ (Table \ref{tab_OPs}).\cite{yip94} Analogous experiments can be used to test for broken reflection symmetries associated with unconventional 1D representations. This idea has been pursued experimentally to test for a $d_{x^2-y^2}$ order parameter in the oxide superconductors.\cite{wol93} Analogous arguments apply for the other residual symmetry groups. Other aspects of the Josephson effect that are specific to unconventional superconductors are discussed in Refs. (\onlinecite{rai87,mil88,yip90,yip93d}).

\vspace*{-3mm}
\subsubsection*{Novel Vortices and Vortex Structures}
\vspace*{-3mm}

The initial discovery of multiple superconducting phases in UPt$_3$ was made in field sweeps of the ultrasound absorption, where a peak was detected at a field of $H\simeq 0.6 H_{c2}$.\cite{mul87,qia87} The existence of such an anomaly immediately suggested the possibility of a structural transition in the flux lattice transition, a vortex-core transition,\cite{sch89} a transition in the background order parameter,\cite{vol88a} or some combination of order parameter transformations. There are a surprising number of possibilities for phase transitions of a two-component order parameter in a magnetic field. Even at the level of a single vortex, there are a number of energetically stable structures. Tokuyasu, {\et}\cite{tok90} investigated vortices in the 2D models for ${\vH}||{\vc}$ and found three classes of stable solutions depending on the material parameters defining the GL functional: (1) an axially symmetric vortex core, (2) a `triangular' vortex core with $C_3$ rotational symmetry and (3) a non-axisymmetric vortex with a reflection rotational symmetry (`crescent vortex'). These vortex structures can be classified by noting that the ground state for the 2D model breaks time-reveresal symmetry, is doubly degenerate and is rotationally symmetric. Assume that the ground state order parameter is ${\veta}_{eq}\sim (1,-i)$ and consider the vortex states in this phase.  The internal structure of these vortices is most easily characterized by the asymptotic form of the vortex order parameter,
\ber
{\veta}(|{\vx}|\gg\xi)
\longrightarrow
&&
\eta_0\,\frac{1}{\sqrt{2}}(1,-i)\,e^{ip\phi}
\nonumber\\
&+&\rho(|{\vx}|)\,\frac{1}{\sqrt{2}}(1,+i)\,e^{im\phi}
\,,
\label{vortex_asymp}
\eer
where $\rho$ decays to zero as $|{\vx}|\rightarrow\infty$, and the integers $p$ and $m$ correspond to the circulation quanta associated with the time-reversed pair of order parameters. Negelecting the SBF, the fourth-order GL functional is invariant under the group of rotations about the ${\vc}$ axis, which simplifies in the classification of the vortex structures. The residual symmetry group of the ground state order parameter is $U(1)_{L_z+I}$, the gauge-rotation group, whose generator is $Q_z=L_z+I$, where $L_z$ is the generator for orbital rotations about ${\vc}$ and $I$ is the generator of gauge transformations. Now ${\veta}_{eq}$ obeys,
\be
Q_z{\veta}_{eq}=0
\,.
\ee
The vortex excitations of this condensate are then classified by the quantum numbers, 
$q$, of $Q_z$;
\be\label{eigenvalue}
Q_z\,{\veta}(|{\vx}|\gg\xi)=q\,\,{\veta}(|{\vx}|\gg\xi)
\,,
\ee
in the asymptotic limit where the anisotropy of the vortex core can be neglected. The compatibility of eq.(\ref{vortex_asymp}) and (\ref{eigenvalue}) places constraints on the quanta of circulation for the two components of the order parameter; $q=p=m+2$. Table~\ref{tab_vortex} lists the lowest vortex quantum numbers, and their identification with the structures calculated in Refs.(\onlinecite{sch89,tok90,tok90a,tok90c,bar91}). Note that asymptotic circulation is given by $p$; the circulation $m$ associated with the time-reversed phase is confined to the core. In addition to the single quantum vortices, there is a doubly-quantized vortex which preserves the axial symmetry ({\ie} $m=0$). This vortex requires no circulation in the time-reversed phase, and as a result has an anomalously low core energy, and is energetically stable compared to two single quantum vortices over a significant region of the GL phase diagram.\cite{tok90a}

\begin{table}
\begin{tabular}{lcc}
Name & p & m \\ 
\hline
Clover   	 & -2 & -4 \\
Triangle	 & -1 & -3 \\
Crescent 	 & +1 & -1 \\
Axial-double	 & +2 &  0 \\
\hline
\end{tabular}
\caption{Quantum numbers for rectilinear vortices ($||\,{\vc}$) in the 2D model}
\label{tab_vortex}
\end{table}

The addition of a SBF introduces an additional aspect to the relative stability of various vortex-core structures which can lead to additional phase transitions.\cite{hes94} At fields above $H_{c1}$ analyses of vortex phases show complex behavior in the vortex lattice structures, including phase transitions between lattices with different symmetry, sometimes driven by an instability in the vortex core order parameter.\cite{tok90a,tok90c,mel92,zhi92} The role of the SBF is important in any GL analysis of the vortex lattice structure, and has recently been investigated by Joynt.\cite{joy91} I will not discuss vortex lattice studies here, but merely emphasize that a local probe of vortex structures, such as STM,\cite{hes89f} would be extremely valuable in sorting out the nature of the vortex structures in UPt$_3$ and providing strong tests for various models of the order parameter.

\vspace*{-3mm}
\subsubsection*{Paramagnetism in Microcrystals and Thin Films}
\vspace*{-3mm}

Paramagnetism can serve as an important probe of the spin structure of the order parameter, particularly as an experimental signature to differentiate even- and odd-parity superconductivity. The effect of a magnetic field, or magnetic surface, is qualitatively different for even- and odd-parity order parameters; in odd-parity, spin-triplet superconductors the transition temperature, energy gap, as well as other properties can depend strongly on the orientation of the field relative to spin-quantization axis of the pairs.\cite{cho91,sun92}

If Cooper pairs form spin-singlets, then the Zeeman energy, which favors an unequal spin population, is pair-breaking for all field directions. In a spin-triplet superconductor the situation is more complicated. Recall that a real ${\vd}$-vector, which in general depends on the relative momenta of the pair, specifies the direction along which the pair $\left({{\vk}_f\;,- {\vk}_f}\right)$ is a pure `opposite spin state', $\vert\uparrow\downarrow\rangle+\vert\downarrow\uparrow\rangle$, {\it i.e.} ${\vd}\left({{\vk}_f}\right)\cdot{\vS}_{pair}=0 $. Conversely, any quantization axis perpendicular to ${\vd}$ is an `equal-spin-pairing' (ESP) direction, with equal amplitudes for the spin projections $\vert\uparrow\uparrow\rangle$ and $\vert\downarrow\downarrow\rangle$. A magnetic field along an ESP direction can easily polarize the pairs (and thus minimize the Zeeman energy) by simply altering the relative number of $\vert\uparrow \uparrow\rangle$ and $\vert\downarrow\downarrow\rangle$ pairs with essentially no loss in condensation energy. Therefore, a magnetic field is not pair-breaking if ${\vH}\perp{\vd}\left({{\vk}_f}\right)$ for all ${\vk}_f$. However, a magnetic field with ${\vH}\vert\vert{\vd}\left({ {\vk}_f}\right)$ is pair-breaking, at least for the pairs $\left({{\vk}_f\;,-{\vk}_f}\right)$, as in the case of conventional spin-singlet pairing.

The spin structure of the order parameter can be probed by measuring the spin susceptibility in the superconducting state. There are significant differences in odd-parity superconductors depending on whether or not there is weak or strong spin-orbit coupling in the pairing channel. For an ESP state, in the absence of spin-orbit coupling the ${\vd}$-vector will orient itself perpendicular to the magnetic field in order to minimize the Zeeman energy.  Thus, the measured spin-susceptibility (for $H\rightarrow 0$) will be unchanged below $T_c$.  However, if there is crystalline anisotropy and strong spin-orbit coupling then a rotation of ${\vd}$ implies an energy cost of order $T_c$. Thus, spin-orbit coupling is expected to select preferred directions for ${\vd}$ in the crystal, and the orientation of the magnetic field may then directly probe the spin structure of the order parameter.

\begin{figure}
\includegraphics[width=\columnwidth]{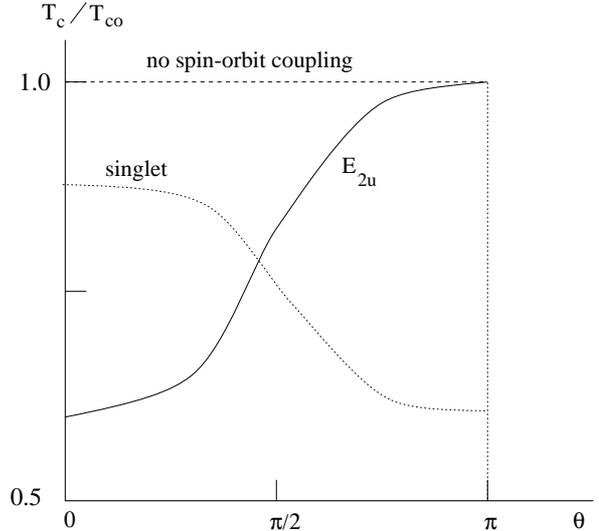}
\caption{Sketch of the transition temperature vs. tilt angle from the ${\vc}$-axis for the three model order parameters: (a) E$_{2u}$ model with ${\vd}||{\vc}$, (b) triplet order parameter with no spin-orbit coupling, and (c) an even parity, spin-singlet order parameter with $\mu_{||}<\mu_{\perp}$. The magnetic pair-breaking parameter is $\frac{\mu_{||}H}{2\pi T_{c0}}= 0.5$.}
\label{fig:paramagnetic_effect}
\end{figure}

The shielding effect of the Meissner current is a complication, which can be avoided by working with crystals that are small compared with the penetration depth. A further complication is surface pair-breaking, which is effectively avoided for crystals of dimension larger than the coherence length. Finally, vortex nucleation is suppressed for dimensions much less than the penetration depth.  Thus, the optimum geometry is a single crystal of characteristic dimensions of order a several coherence lengths, $\xi\ll t\ll\lambda$. In this limit the shielding current can be ingnored, and the order parameter is approximately uniform over the sample. The transition to the superconducting state in a field is then determined by the Zeeman coupling. Fig.~\ref{fig:paramagnetic_effect} shows the transition temperature as a function of tilt angle $\vartheta$ of the applied field ${\vH}$ relative to the ${\vc}$ axis of UPt$_3$, for three models: (1) an odd-parity order parameter with ${\vd}||{\vc}$ and strong spin-orbit coupling, (2) an odd-parity order parameter with no spin-orbit coupling, and (3) an even-parity order parameter and $\mu_{||}<\mu_{\perp}$.  Note that the model of Ref. (\onlinecite{mac91}) predicts a nearly isotropic transition temperature because the ${\vd}$-vector is free to rotate perpendicular to the field, or more precisely, for fields $\mu H\sim T_c$ the anisotropy energy associated with SBF is small compared to the characteristic Zeeman energy, which favors ${\vd}\perp{\vH}$.

\vspace*{-3mm}
\subsubsection*{Collective Modes: \\ Circular Birefringence and Broken ${\cal T}$-symmetry}
\vspace*{-3mm}

One of the principal techniques for investigating the symmetry and low-lying collective excitations of the order parameter in superfluid $^3$He is high-frequency longitudinal and transverse ultrasonics.\cite{hal90} Resonances between acoustic modes and collective modes lead to sharp features in the frequency and temperature dependence of absorption and velocity. A similar spectroscopy of collective excitations using high-frequency EM probes has been investigated theoretically for several unconventional order parameters.\cite{hir89,hir92} The observation of an order parameter collective mode in UPt$_3$ would be an important experiment; it would provide direct evidence of a multi-component order parameter and could possibly be used to determine additional information on the residual symmetry group ({\it cf.} related studies in superfluid $^3$He\cite{mck90}) and test for specific broken symmetries, {\eg} broken time-reversal symmetry and broken reflection symmetries.\cite{yip92} All of the candidates for the order parameter listed in Table \ref{tab_OPs} for the low-temperature phase of UPt$_3$ break time-reversal (${\cal T}$) symmetry. It is important to test this prediction experimentally.

One possibility for detecting broken ${\cal T}$-symmetry would be to observe circular birefringence (CB) and/or dichroism (CD) in the reflectivity of electromagnetic or transverse acoustic waves with ${\vq}||{\vc}$ (see Ref.(\onlinecite{yip92}) and references therein). This polarization effect is present if the ground state exhibits broken reflection symmetry, broken time-reversal symmetry and broken particle-hole symmetry, conditions satisfied by the ground state $E_{2u}$ order parameter as well as other candidates. The effect is small; the nominal magnitude of the elliptical polarization expected for CB is of order $\delta\theta_{CB}\sim(v_f/c)(\xi/\lambda) (\Delta/E_f)\ln(E_f/\Delta)\lesssim 1\,\mu{\rm rad}$. However, the CD/CB signal originates from the asymmetry in the coupling to a collective mode of the order parameter (due to the internal orbital currents) for right- vs. left-circularly polarized waves. As a result, the CD/CB signal is enhanced for frequencies near the resonance frequency of the collective mode, typically
$\omega\sim\Delta$.\cite{yip92}

Muon spin-relaxation measurements on UPt$_3$ have recently been reported\cite{luk93} which indicate broken ${\cal T}$-symmetry below the second superconducting transition ($T<T_{c*}$) with an internal field of order $0.1\,G$. At present it is not known if the increase in the $\mu$SR relaxation is due to the broken ${\cal T}$-symmetry of the superconducting order parameter; however, small internal fields of this magnitude are characteristic of the orbital currents associated with spontaneous ${\cal T}$-violation in the low-temperature phase of the orbital 2D models.  Observable measures of the broken time-reversal symmetry are expected to be small because these typically effects vanish to leading order in $T_c/E_f$, at least in the clean limit as is clear from eqs.  (\ref{f_grad}) and (\ref{kappas}). The leading contribution to $\kappa_2-\kappa_3$ comes from particle-hole asymmetry which is nominally of order $\kappa_2-\kappa_3\sim\frac{T_c}{E_f}\kappa_1$. This leads to an orbital magnetic moment that is of order $M_{orb}\sim H_{c1}(T_c/E_f)/\ln(\lambda/\xi)\sim 1\,G$. In an ideal, bulk material this field will be completely screened by surface currents. However, inhomogeneities, for example polycrystals with dimensions $\lesssim \lambda$, impurities and vortices all inhibit perfect screening of the internal field. Choi and Muzikar\cite{cho89b} calculated the internal field induced by a non-magnetic impurity in a superconductor with an orbital ground state that breaks ${\cal T}$-symmetry; their result is $B(0)\simeq \frac{\pi^2}{18}(\frac{e}{c})\sigma_{imp}(\xi_0/a)(N_f \Delta^2)$, where $\sigma_{imp}=\pi a^2$ is the scattering cross section of the impurity and $N_f\Delta^2$ is the condensation energy density. Within a factor of ${\cal O}(1)$ the magnitude of the field is $B(0)\simeq(a/\xi_0)H_{c1}(0)/\ln(\lambda/\xi)\sim 0.1-1.0\, G$ for reasonable estimates of the impurity scattering radius.

Broken ${\cal T}$-symmetry by the superconducting order parameter could have dramatic effects on flux penetration and flux motion in UPt$_3$.  In the orbital 2D models the broken ${\cal T}$-symmetry ground state is doubly degenerate reflecting the two possible orientations ($\pm{\vc}$) of the orbital moment. Internal orbital pair currents are expected to generate an asymmetry in $H_{c1}$ for the nucleation of vortices parallel or anti-parallel to the orbital moment.\cite{tok90} The asymmetry of $H_{c1}^{\pm}$ for ${\vH}||\pm{\vc}$ measures the difference in energy associated with core structures (Table~\ref{tab_vortex}) of vortices with equal magnitude, but opposite sign, circulation (or supercurrent ${\vv}_s\sim{\vpartial}\vartheta+\frac{2e}{\hbar c}{\vA}$) relative to the {\it internal} orbital current. The magnitude of $(H_{c1}^{+}-H_{c1}^{-})$ reflects both the spontaneous breaking of ${\cal T}$-symmetry by the bulk order parameter and the spontaneous breaking rotational symmetry in the vortex core; particle-hole symmetry, as measured by $\kappa_2-\kappa_3$, is not required for an asymmetry of $H_{c1}^{\pm}$, although it enhances the efffect.\cite{tok90}

The same two-fold degeneracy that is responsible for an asymmetry of $H_{c1}^{\pm}$ provides a mechanism for masking it. Domain walls separating regions of oppositely directed orbital moment are likely to develop when the material is cooled below the second transition temperature, unless specific conditions are taken to prepare a single superconducting domain. Like vortices, domain walls are regions in which the order parameter is strongly deformed, and are expected to be metastable, particularly if there are structural defects present to pin the domain walls. If domain walls are present then the asymmetry in $H_{c1}^{\pm}$ will likely be unobservable since vortices will enter the domains with the smaller $H_{c1}$, or flux will be channeled by the domain wall itself. This latter possibility was suggested by Sigrist, {\et}\cite{sig89}, who examined the magnetic structure of vortices bound to a domain wall and estimated the nucleation energy for a single vortex at a domain wall to be lower than the nucleation energy of a vortex in bulk.

Recent measurements of the decay of remnant flux in single-crystals of UPt$_3$ exhibit non-thermally activated flux creep, with time-scales of order $10^4 - 10^5$ seconds for temperatures ranging from $7\,,mK-350\,mK$.\cite{mot94} Flux creep due to quantum tunneling of vortices predicts a decay rate $d\ln M/d\ln t\simeq -Q_u(j_c/j_0)^{1/2}$; $j_0$ is the  depairing current density, $j_c$ is the critical current density, and $Q_u=(e^2/\hbar)(\rho_n/\xi)$, where $\rho_n$ is the normal-state resistivity and $\xi$ is the coherence length for $T\rightarrow 0$.\cite{bla94} The experimental data on flux creep in UPt$_3$ is qualitatively different; the decay does not follow a logarithmic behavior, and the creep rate is much faster than expected from quantum tunneling of vortex bundles.\cite{mot94} The remanant magnetization appears to decay in two steps; slow initial decay over timescales of $10^2\,secs$, followed by a fast decay from $10^2-10^4\, secs$, which appears to be roughly temperature independent below $T\simeq 120 \, mK$. The experiments suggest that there is more to the mechanism of the flux decay in UPt$_3$ than quantum tunneling of vortex lines.

The internal structure of planar, superconducting domain walls separating broken ${\cal T}$-symmetry phases is described by an order parameter with the approximate form, ${\veta}=\eta_0(\cosh(x/\xi)\,,\,i\sinh(x/\xi))$, where $\xi$ is of order the coherence length and $x$ is the coordinate perpendicular to the domain wall. Supercurrents flow along the domain wall and generate a local field which reverses sign across the wall.\cite{gor87,sig89} In an external field a difference in the population of vortices in the regions of lower and higher $H_{c1}$ will exert a net force on the domain wall. Because of domain wall motion, pinning of domain walls by defects and the possibility of vortex channeling along the walls, the flux flow properties of UPt$_3$ with a broken ${\cal T}$-symmetry phase at low temperature are expected to show qualitative differences compared to conventional type II superconductors.

\vspace*{-3mm}
\subsection*{Conclusion}
\vspace*{-3mm}

In summary, the detailed measurements of the phase diagram, combined with the pressure-dependent correlation between the AFM order and the double transition in zero field, provide strong constraints on the symmetry and dimensionality of the order parmeter for UPt$_3$. The low-temperature anisotropy of the upper critical field is interpreted in terms of an odd-parity, spin-triplet state with ${\vd}||{\vc}$ enforced by strong spin-orbit coupling. This interpretation appears to conflict with the model of Machida and Ozaki based on a spin-triplet order parameter and effectively no spin-orbit coupling. The phase diagram, including an apparent tetracritical point for all field orientations, can be explained naturally within the 2D E$_{2u}$ model provided the hexagonal anisotropy is weak, which is consistent with absence of in-plane hexagonal anisotropy of $H_{c2}(T)$ at low temperatures.

Although the E$_{2u}$ order parameter seems to be able to explain a number of basic features of the superconducting phases of UPt$_3$, there are many important open questions. For example: (i) What is the pairing mechanism and the origin of the correlation between the basal plane AFM order and superconductivity?, (ii) Is UPt$_3$ unique (and if so why) among the U-based heavy fermions in exhibiting multiple superconducting phases in its pure, stochiometric phase?, (iii) What is the residual symmetry group and the detailed structure of $\Delta({\vk}_f)$?, and so on.

Unfortunately, what is not within easy reach of existing theory is the pairing {\it mechanism}. Even if one accepts the spin-fluctuation-exchange model as a reasonable starting point for estimating the pairing interaction in UPt$_3$, different RPA-type models for the pairing interaction, with information on the dynamic spin susceptibility obtained from neutron scattering data as an input, give different predictions for the symmetry channel.\cite{nor87,put88,nor90,nor91} Even the parity of the dominant pairing channel depends on additional assumptions about the input parameters to the effective interaction.\cite{nor91} And if phenomenological approaches to the effective interaction converge to a robust solution, the outstanding problem remains: to develop a systematic procedure, presumably based on some small expansion parameter, for identifying the dominant contributions to the pairing interaction. To date no such theory exists for the heavy fermion superconductors. Given that the basic mechanism of pairing is not understood, any microscopic model of the coupling of superconductivity to a weak SBF will suffer from this uncertainty.

Fortunately, Fermi-liquid theory appears to work well for UPt$_3$ (and some of the other heavy fermion superconductors) so one can be optimistic that a solution to the identity of the phases of UPt$_3$, and presumably other heavy fermion superconductors, is within reach.  A number of key experiments, some discussed above, will hopefully be carried out in order to provide more direct tests of the residual symmetry group of the order parameter, as well as the SBF model for UPt$_3$. Many of the remaining problems and interpretation of these experiments based on of various models will require the full power of the Fermi-liquid theory of superconductivity.

\vspace*{-3mm}
\subsection*{Acknowledgements}
\vspace*{-3mm}

I thank Anupam Garg and Ding Chuan Chen for discussions on the interpretation of the phase diagram. I especially want to thank my collaborators, Chi-Hoon Choi, Daryl Hess, Paul Muzikar, Dierk Rainer, Taku Tokuyasu, Sungkit Yip and Dandong Xu for many discussions on superconductivity in heavy fermions. I am indebted to my experimental colleagues, Shireen Adenwalla, Bill Halperin, John Ketterson, Moises Levy, Shiwan Lin, Mark Meisel, B. Shivaram and especially Bimal Sarma for discussions and their willingness to share their results with me.  This work was supported in part by the NSF (grant DMR-9120521) though the Materials Research Center at Northwestern University, and by the Nordic Institute for Theoretical Physics (NORDITA). I also acknowledge the support of the Aspen Center for Physics where a draft of this manuscript was prepared.


\end{document}